\documentclass[prb,twocolumn,showpacs,preprintnumbers,superscriptaddress,amsmath,amssymb,10pt]{revtex4-1}

\usepackage[pdftex]{graphicx}

\usepackage{multirow}

\usepackage{dcolumn}

\usepackage{color}
\definecolor{rot}{rgb}{1,0,0}

\newcommand{\blau}{}   

\usepackage[justification=raggedright,singlelinecheck=false]{caption}

\begin{document}

\title{Interlayer excitonic spectra of vertically stacked MoSe$_2$/WSe$_2$ heterobilayers}
\author{Roland Gillen}\email{roland.gillen@fau.de}
\affiliation{Department of Physics, Friedrich-Alexander University Erlangen-N\"{u}rnberg, Staudtstr. 7, 91058 Erlangen, Germany}

\date{\today}

\begin{abstract}
The optical spectra of vertically stacked MoSe$_2$/WSe$_2$ heterostructures contain additional ’interlayer’ excitonic peaks that are absent in the individual monolayer materials and exhibit a significant spatial charge separation in out-of-plane direction. Extending on a previous study, we used a many-body perturbation theory approach to simulate and analyse the excitonic spectra of MoSe$_2$/WSe$_2$ heterobilayers with three stacking orders, considering both momentum-direct and momentum-indirect excitons. We find that the small oscillator strengths and corresponding optical responses of the interlayer excitons are significantly stacking-dependent and give rise to high radiative lifetimes in the range of 5-200\,ns (at T=4\,K) for the 'bright' interlayer excitons. Solving the finite-momentum Bethe-Salpeter Equation, we predict that the lowest-energy excitation should be an indirect exciton over the fundamental indirect band gap (K$\rightarrow$Q), with a binding energy of 220\,meV. However, in agreement with recent magneto-optics experiments and previous theoretical studies, our simulations of the effective excitonic Land\'e g-factors suggest that the low-energy momentum-indirect excitons are not experimentally observed for MoSe$_2$/WSe$_2$ heterostructures. We further reveal the existence of 'interlayer' C excitons with significant exciton binding energies and optical oscillator strengths, which are analogous to the prominent band nesting excitons in mono- and few-layer transition-metal dichalcogenides. 
\end{abstract}

\maketitle

\section{Introduction}
The experimental realization of Graphene in 2004~\citep{graphene-2004,novoselov-2005} inspired an ongoing search for novel energetically stable quasi-two-dimensional materials and a detailed experimental and theoretical study of the influence of their reduced dimensionality. Recent theoretical reports based on high-throughput screening and machine-learning methods suggest that several thousand exfoliable, and largely unsynthesized, materials exist~\cite{new-materials-1,new-materials-2}, offering a rich pool of materials with diverse physical properties.

A particularly interesting and well-researched family of novel two-dimensional materials are the transition metal dichalcogenides (TMDC) of molybdenum and tungsten with the structural formula $MX_2$ (with $M$=Mo,W and $X$=S,Se,Te). These materials have been well known in their bulk phases, where they are indirect semiconductors with band gaps in the range of 0.8-1.3\,eV and assume a layered structure with hexagonal symmetry. Analogously to graphite, the atoms within the layers are bonded covalently, while the layers are bound together mainly through non-covalent interactions. Interestingly, it had been realized early that this interlayer interaction, albeit weak {\blau compared to the covalent intralayer bonds}, has a significant effect on the electronic and optical properties of molybdenum and tungsten TMDCs: 
the fundamental band gap is indirect in bulk and few-layer samples and increases with decreasing thickness due to quantum confinement effects and the attenuation of interlayer-coupling induced splittings of the valence and conduction band edges. In the absence of interlayer coupling effects in the monolayer limit, the band gap is \textit{direct}, causing a strong enhancement of photoluminescence quantum yield for decreasing material thickness~\citep{mak-2010,splendiani-2010,MoSe2WSe2MoS2-PL,WS2-PL,MoS2-tunable-PL,scheuschner-2014}. 
\begin{figure*}
\centering
\includegraphics*[width=0.99\textwidth]{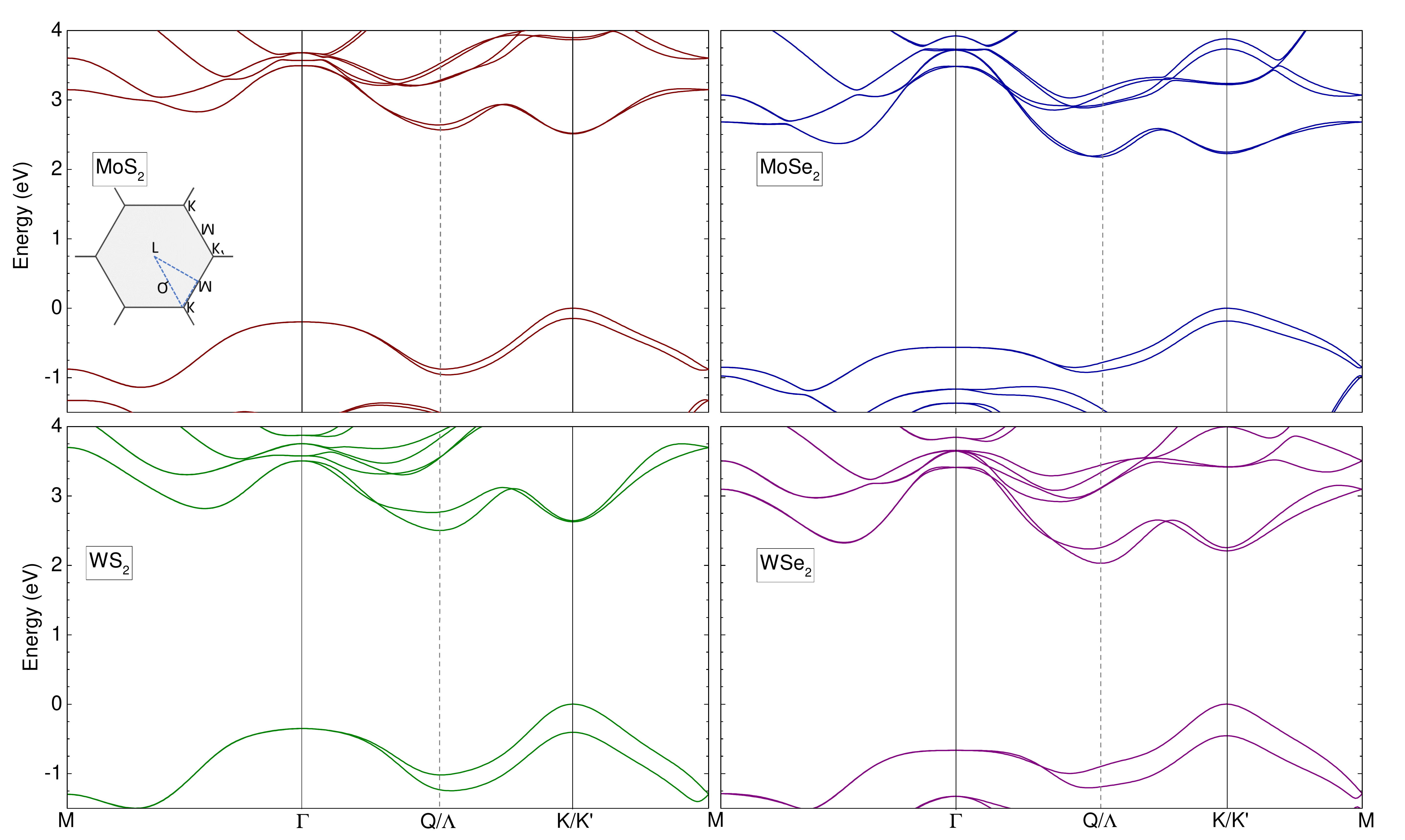}
\caption{\label{fig:monolayer-bands} (Color online) Calculated electronic dispersion of four monolayer molybdenum and tungsten dichalcogenides from simulations on the G$_0$W$_0$ level of theory~\cite{footnote1}. {\blau The conduction band valley about half-way along the $\Gamma$-K/K' lines is typically labeled $Q$ or by the greek index $\Lambda$.}
}
\end{figure*}

At the same time, the reduced dimensionality and highly {\blau non-local} dielectric screening of Coulomb interaction give rise to rather large exciton binding energies on the order of {\blau 0.3-0.6}\,eV for monolayer molybdenum {\blau and tungsten TMDCs~\cite{MoS2-exciton-binding-2,WSe2-binding1,ugeda-MoSe2-excitons,WS2-binding1,MoS2-WS2-exciton-binding-hill,WSe2-binding2,MoS2-WS2-exciton-binding-rigosi,chernikov-review,Goryca2019}} despite the exciton wavefunctions to be rather expanded, with typical Bohr radii on the order of {2}\,nm~\cite{Goryca2019,qiu-2013,exciton-paper}. Further, the electronic band structures of molybdenum and tungsten disulfides, -selenides and -tellurites show several local conduction and valence band extrema of similar energies, see Figure~\ref{fig:monolayer-bands}. This gives rise to a rich excitonic spectrum in mono- and few-layer TMDCs, dominated by 'A' excitons originating from the direct band gaps and, in few-layer materials, 'I' transistions associated with the indirect fundamental band gap. Interestingly, strongly bound 'C' excitons appear at higher energies, between 2.0 and 3.0\, eV, depending on the material, arising from a band nesting between valence and conduction band edges and an associated high-joint density at special points in the Brillouin zone~\cite{carvalho-band-nesting,qiu-2013,exciton-paper}. 

The strong spin-orbit interaction introduced by the heavy transition metal atoms further leads to a prominent splitting of the valence band maximum at the corners of the two-dimensional hexagonal Brillouin zones, allowing to exploit the circular dichroism for valley selective optical excitations with circularly polarized light. 
These findings inspired the use of mono- and few-layer TMDCs for a range of novel applications of TMDCs in novel thin and flexible optoelectronic devices, such as photodiodes\cite{led1,led2}, photodetectors\cite{photodetector1} or single-photon emitters\cite{single-photon2,single-photon3,single-photon4,single-photon5}, {\blau and} for a combination of spin- and valleytronics\cite{spinvalley-1, spinvalley-2}.

A possible way to further tailor the electronic and optical properties of TMDCs is the combination of different TMDCs to form vertically stacked heterostructures. Here, the non-covalent interlayer interaction in principle allows for atomically sharp and essentially strain-free interfaces even for lattice-mismatched materials; particularly interesting compared to conventional heterostructures formed of 3D bulk materials, where interfacial defects can have a significant effect on the material properties.
Recently, the combination of Mo and W based TMDCs arose substantial interest: for MoS$_2$/WS$_2$\cite{MoS2WS2-1}, MoS$_2$/WSe$_2$\cite{MoS2-WSe2-1,kunstmann-interlayer-excitons,Karni-Mos2Wse2}, MoSe$_2$/WSe$_2$\cite{WSe2MoSe2-2,WSe2MoSe2-3,Nayak-MoSe2WSe2-twist,moire-MoSe2WSe2,geim-MoSe2WSe2} and MoSe$_2$/WS$_2$\cite{MoSe2WS2-1,MoSe2WS2-2} heterostructures, experimental observations indicate the presence of additional photo\-luminescence (PL) signals from excitations with long lifetimes of 1-100\,ns~\cite{WSe2MoSe2-2,miller-MoSe2WSe2}, which are absent in the PL spectra of the individual monolayer materials.

The detailed origin of these transitions is still not entirely clear and might vary with the material combination {\blau and experimentally studied samples}. {\blau Theoretical studies} predict that the aforementioned heterostructures form type-II heterostructures with the valence and conduction band edges of the composite materials being localized in different layers~\cite{kang-offsets}. {\blau For this reason, a reasonable and popular explanation of the observed signals is} the attribution to interlayer excitons with a distinct spatial separation of the bound electrons and holes, an interesting concept for application in photovoltaics based on such heterostructures~\cite{MoSe2-WSe2-photovoltaics}. These arguments are based on the assumption that the interlayer coupling between the individual materials is small enough that the electronic properties of the individual materials are largely unaffected. However, it is known from the TMDC \textit{homo}-multilayer materials, such as bilayer MoS$_2$, that the (rather weak) hybridization between S or Se $p_z$ states in the different layers causes an interlayer-induced band splitting and a layer-number dependent direct-to-indirect band gap transition. 
{At the same time}, these additional signals are somewhat reminiscent of the 'I' transitions in TMDC homo-bilayers and might indicate phonon- or defect-assisted emission over an indirect band gap. 

Due to the rather complex nature of the TMDC band structures with several conduction band and valence band valleys with similar energies and additional spin-orbit and interlayer coupling effects, a clear identification of the origin of these peaks is non-trivial and strongly depends on a comparison of experimental data with theoretical predictions of the electronic bandstructure, exciton binding energies and the orbital composition of relevant bands. For instance, Kunstmann~\emph{et al.} used a correlation of the PL spectra in MoS$_2$/WSe$_2$ heterostructures with various twist angles with the respective DFT band gaps to attribute the prominent 'interlayer' PL peak at around 1.6\,eV to a momentum-indirect exciton, with the bound electron and hole being located at the K point and the center of the hexagonal Brillouin zone, respectively~\cite{kunstmann-interlayer-excitons}. {\blau Additionally}, recent PL measurements on BN-encapsulated MoS$_2$/WSe$_2$ heterostructures with small twist angle suggest the presence of an additional excitonic signal at an energy of about 1\,eV, well below the fundamental band gap of the heterostructure, with a significant Stark shift indicating a strong vertical separation of the electron-hole pair by about 6\,\AA. Based on theoretical simulations, this infrared peak was attributed to a transition between the global valence band maximum at the K point of the WSe$_2$ layer and the global conduction band minimum at the K-point of the MoS$_2$ layer~\cite{Karni-Mos2Wse2}. 

For nearly lattice-matched MoSe$_2$/WSe$_2$ heterostructures, low-temperature experiments revealed that the prominent interlayer peak consists of a doublet of two separate contributions with a small energy difference of 25\,meV. Due to the aforementioned complexities of the electronic band structure, the exact origin or these peaks is still a matter of debate. Depending on the experimental evidence and insight from theoretical calculations, the PL signals were attributed to (i) a mix of momentum direct ($K\rightarrow K$) and momentum-indirect ($K\rightarrow Q$) interlayer excitons, evidenced by the different temperature behaviors of the PL peak lifetimes~\cite{WSe2MoSe2-2,miller-MoSe2WSe2}, (ii) to a pair of momentum-indirect $K\rightarrow Q$ excitons, with an energy separation due to the spin-orbit splitting of the conduction band {\blau valley} at the $Q$ point, evidenced by the observed opposite circular polarization~\cite{hanbicki-MoSe2WSe2}, (iii) a singlet-triplet pair of neutral, momentum-direct, excitons at the K and K' points, evidenced by recent measurements of the {\blau interlayer exciton photoluminescence under application of external magnetic fields}~\cite{Ciarrocchi-nature-MoSe2WSe2,Wang2020,Delhomme2020,foerg2020moire}, and (iv) a pair of spatially indirect neutral exciton and spatially indirect negatively charged trion, evidenced by an observed constant relative intensity of the two PL signals at low temperature~\cite{geim-MoSe2WSe2}, but somewhat contradicted by the circular polarization response reported by other authors. For structurally similar MoS$_2$/WS$_2$ hetero-bilayers, experimental reports suggest the existence of three interlayer exciton peaks, which were attributed to $K\rightarrow K$, $Q\rightarrow\Gamma${\blau,} and $K\rightarrow\Gamma$ transitions based on a combination of optical measurements and DFT+BSE~\cite{Okada-MoS2WS2}.

An alternative explanation for the origin of the observed interlayer exciton peak structure is offered by the effects of the intrinsic (albeit small) lattice mismatch and small deviations in relative twist angles from the symmetric stacking orders in studied samples, giving rise to long-period {\blau moir\'e } superlattices and corresponding {\blau moir\'e } potentials, which might activate otherwise optically forbidden spin-flip transitions~\cite{Ciarrocchi-nature-MoSe2WSe2} and allow for a {\blau moir\'e } trapping and localization of excitonic states in the minima of the {\blau moir\'e } superlattice potential~\cite{Seyler-MoSe2WSe2-moire}, of significant potential interest for manipulation of excitons in van-der-Waals heterostructures. Additionally, the differences in local stacking order have been shown to affect the interlayer exciton binding energies, oscillator strengths{\blau,} and optical selection rules~\cite{gillen-interlayer}. Based on magneto-luminescence experiments, a recent study suggested that the interlayer PL peaks in twisted MoSe$_2$/WSe$_2$ heterobi- and trilayer systems in R registry arise from {\blau moir\'e } potential modulated momentum-direct and phonon-assisted momentum-indirect interlayer exciton emissions, respectively~\cite{Hoegele-MoSe2WSe2-moire}. 

On the theoretical side, accurate \textit{\blau ab initio} investigations of the excitonic spectra of van-der-Waals hetero\-structures are limited to rather small and symmetric systems due to the delicate interplay of quasiparticle bandstructure, effects from interlayer interaction and spin-orbit coupling and (computationally expensive) Coulomb interaction between electrons and holes. Despite these constraints, a number of studies using various methods have been reported recently, typically focussing on the properties of momentum-direct interlayer excitons~\cite{palumno-2015,lantini-2017,gillen-interlayer,wirtz-interlayer,Okada-MoS2WS2,deilmann-MoS2WS2,interlayer-moire-bse} and trions~\cite{deilmann-MoS2WS2}. On the other hand, to the best of our knowledge, there is no reliable theoretical data so far on the binding energies and wavefunctions of momentum-indirect excitons, which could significantly contribute to the detailed understanding of the experimental observations of interlayer exciton emissions. Another question that has not yet been addressed  concerns the structure of the absorption above the intralayer band gaps, which is dominated by 'C' excitons with high binding energy and significant interlayer delocalization in TMDC homo-single- and -fewlayer systems~\cite{exciton-paper}. These 'C' excitons have been found recently to lead to interesting resonant enhancement effects in Raman spectroscopy experiments on TMDC multilayer systems~\cite{scheuschner-interlayer-modes,witz-raman}. This raises the questions of whether similar excitons exist in TMDC heterostructures as well and how are they affected by changes in the electronic structure.

The paper is structured as follows: In the first part, the excitonic spectra of molybdenum and tungsten based monolayer TMDCs will be introduced as a basis for the following discussion. {\blau In the second part, using} the example of MoSe$_2$/WSe$_2$ heterobilayers{\blau,} we will then {\blau review} electronic structure and absorption spectra for different stacking orders as obtained from state-of-the{\blau-art} many-body perturbation theory calculations, which yield an accurate prediction of electronic band gaps and exciton binding energies and wavefunctions, thus allowing for a direct comparison with experiment. In the third part, we will discuss the binding energies and wavefunctions of momentum-\textit{indirect} excitons that might contribute {\blau to the experimentally measured optical spectra}. In particular, our calculations suggest that the observed interlayer exciton doublet is unlikely to consist of a combination of momentum-direct and momentum-indirect excitons, in agreement with recent experimental evidence, and point towards an identification of the interlayer exciton doublet with the $K\rightarrow Q$ transitions. In the fourth part, we will show that the similarities in the electronic structure between MoSe$_2$ and WSe$_2$ monolayers should lead to the occurrence of 'interlayer' C excitons in the excitonic spectra of MoSe$_2$/WSe$_2$ heterobilayers, with a significant degree of spatial separation between the bound electrons and holes and rather high exciton binding energies of 400\,meV.
We refer to Refs.~[\onlinecite{exciton-paper}] {\blau and~[\onlinecite{gillen-interlayer}] for the theoretical details, unless indicated otherwise in the text.}
\begin{figure*}
\centering
\includegraphics*[width=0.99\textwidth]{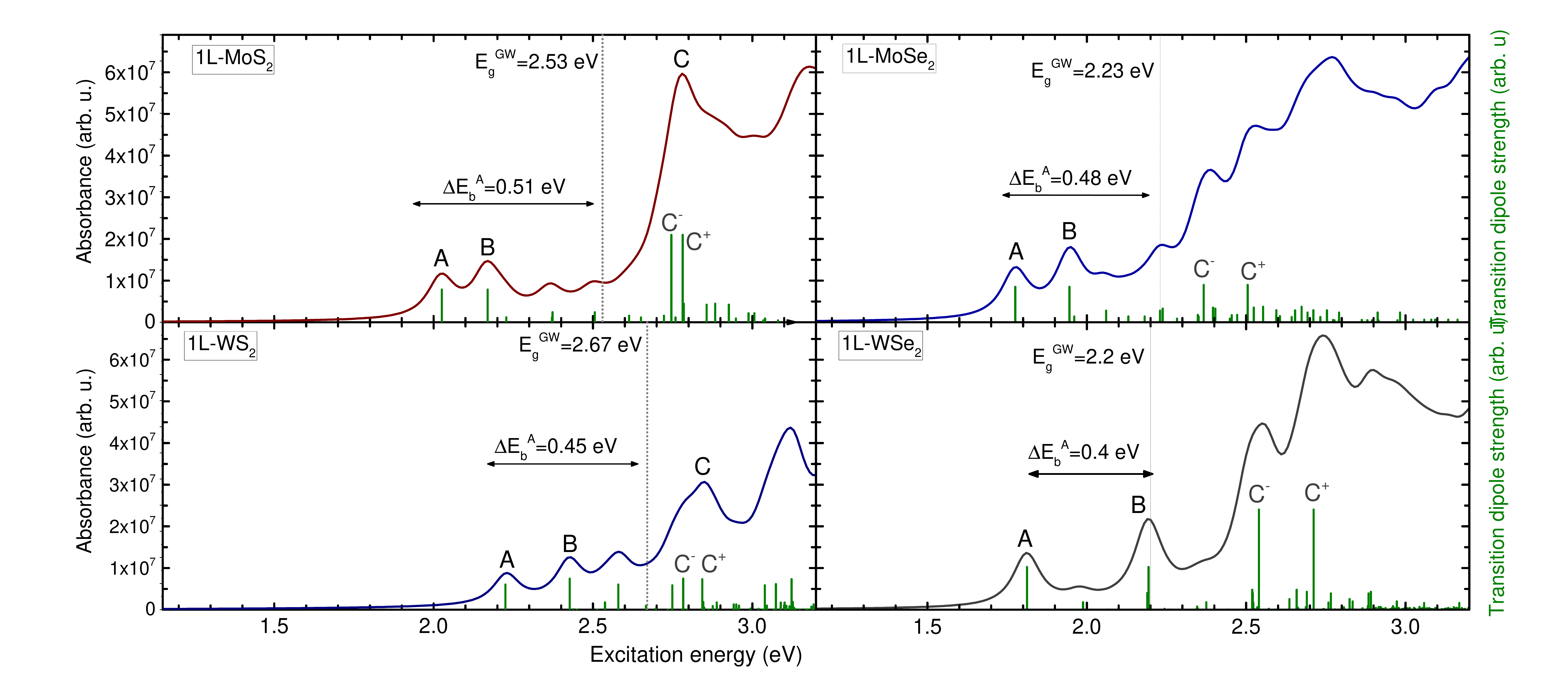}
\caption{\label{fig:monolayer-abs} (Color online) Calculated imaginary parts of the dielectric functions of four materials, including electron-hole coupling effects. The MoS$_2$ and MoSe$_2$ are adapted with permission from Ref.~{\blau[\onlinecite{exciton-paper}]}, the absorption spectra were shifted to match the electronic band gap with the corresponding direct gaps in Figure~\ref{fig:monolayer-bands}. The absorption spectra of WS$_2$ and WSe$_2$ were calculated based on the electronic structure shown in Figure~\ref{fig:monolayer-bands}, following the method outlined in Ref.~{\blau[\onlinecite{exciton-paper}]}. The absolute peak positions are slightly overestimated compared to experiment due to the neglect of temperature effects, which, for instance, can cause a red-shift of the predicted A and B peak positions in MoS$_2$ by about 0.1\,eV~\citep{mol-sanchez-exciton-temperature}. }
\end{figure*}

\section{Results and Discussion}
\subsection{Excitonic spectra of monolayer TMDCs}
\begin{figure}
\centering
\includegraphics*[width=0.99\columnwidth]{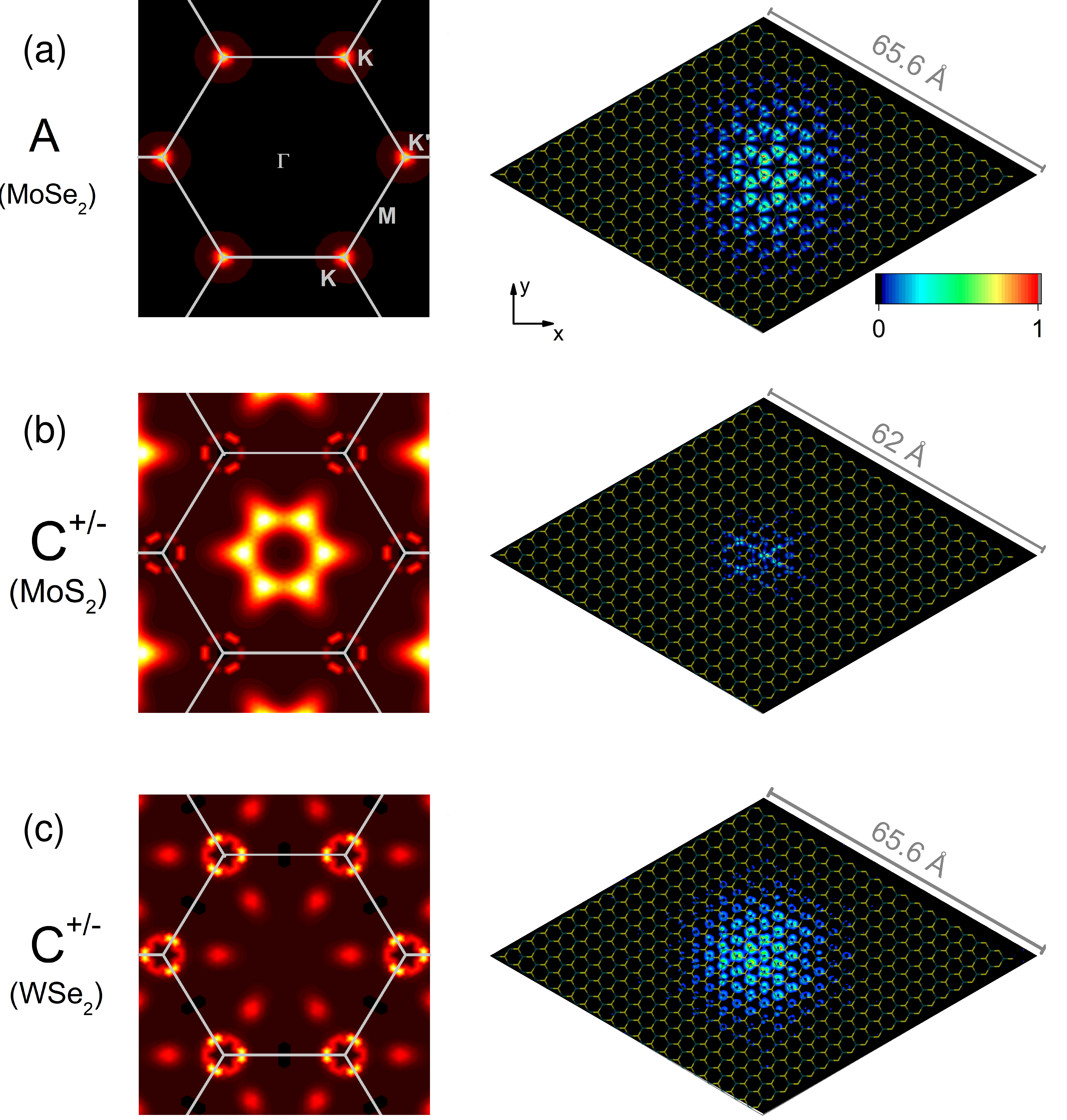}
\caption{\label{fig:exciton-wvfns} (Color online) Calculated exciton wavefunctions of selected \textsl{A} and \textsl{C} excitons. The left side of the subfigures shows the k-resolved contributions of transitions to the exciton wavefunctions. The right side shows top views of the spatial extent of the {\blau electron part} of the exciton wavefunctions. For all excitons, the hole was fixed at a transition-metal atom in the center of the supercell of 20x20 unit cells that was used for the plot (arrows). Adapted with permission from Ref.~{\blau[\onlinecite{exciton-paper}]}. }
\end{figure}

In contrast to common bulk semiconductor materials, the {\blau non-local} dielectric screening {\blau of the Coulomb interaction} in two-dimensional materials introduces a significant Coulomb interaction between optically excited electron-hole pairs{\blau~\cite{WS2-exciton-binding-1}}, necessitating an explicit inclusion of electron-hole coupling effects in calculations not only for a proper description of oscillator strengths of optical excitations{\blau,} but also for accurate exciton binding energies.

Figure~\ref{fig:monolayer-abs} shows the imaginary parts of simulated dielectric functions of the widely researched monolayer molybdenum- and tungsten-based disulfides and diselenides~\cite{exciton-paper}, in good qualitative and quantitative agreement with recent theoretical reports by other groups and consistent with experimentally measured absorption spectra~\cite{MoS2-absorption,TMDC-absorption}.
For all four considered materials, the Coulomb attraction is sufficiently strong to pull a number of prominent absorption peaks well below the {\blau single}-particle absorption onset given by the direct electronic band gaps. An analysis of the calculated excitonic wavefunctions, as shown for MoSe$_2$ in Figure~\ref{fig:exciton-wvfns}~(a), suggests that the lower-energy 'A' peak originates from spin-conserving transitions over the direct band gaps at the K and K' points in the hexagonal Brillouin zone. With this information, one can directly calculate the binding energy of the 'A' electron-hole pairs as the difference between the exciton peak position (i.e. the \textit{optical} band gap) and the corresponding electronic band gap. Here, a minor complication arises for tungsten-based materials, where the spin-order of the spin-orbit split conduction band minimum at the K and K' points is reversed compared to molybdenum-based TMDCs; the lowest-energy direct transition between the valence band maximum is optically dark due to violation of spin-conservation while the transition to the second-lowest conduction band is optically bright~\cite{andor-paper}.
With these considerations in mind, recent theoretical calculations yielded binding energies in the range of 0.4-0.5\,eV for the four considered monolayer materials~\cite{qiu-erratum,wirtz-mos2-excitons,gunnar-MoS2,komsa-excitons,compmatdatabase,Berkelbach-binding,exciton-paper}, well within the range of {\blau 0.3-0.5}\,eV reported from experiments{\blau ~\cite{chernikov-review,Goryca2019}}. 

The large excitonic binding energies are mirrored in the significant spatial localization of the excitonic wavefunctions compared to typical length scales of Mott excitons in 3D semiconductor materials; we find Bohr radii on the order of about {\blau 2}\,nm for the 'A' excitons in all four materials [refer to right side Figure~\ref{fig:exciton-wvfns}~(a)]. The 'B' peak arises from an excitonic transition between the energetically lower sub{\blau-}band of the spin-orbit split valence band maximum at the K/K' points to the spin-matched sub{\blau-}band of the spin-orbit split conduction band minimum, thus appearing at slightly higher energies (mainly determined by the value of the spin-orbit splitting of the valence band), but with similar excitonic wavefunctions and exciton binding energies. The valence and conduction band extrema are completely spin-polarized at the K and K' points, with the electron spin pointing in out-of-plane direction and the spin-order at the three K' points of the hexagonal Brillouin zone reversed compared to the K points~\cite{andor-paper}, thus allowing for a valley-selective excitation of the A and B excitons using circularly polarized light~\cite{TMDC-valley-pol}.
\begin{figure}
\centering
\includegraphics*[width=0.49\columnwidth]{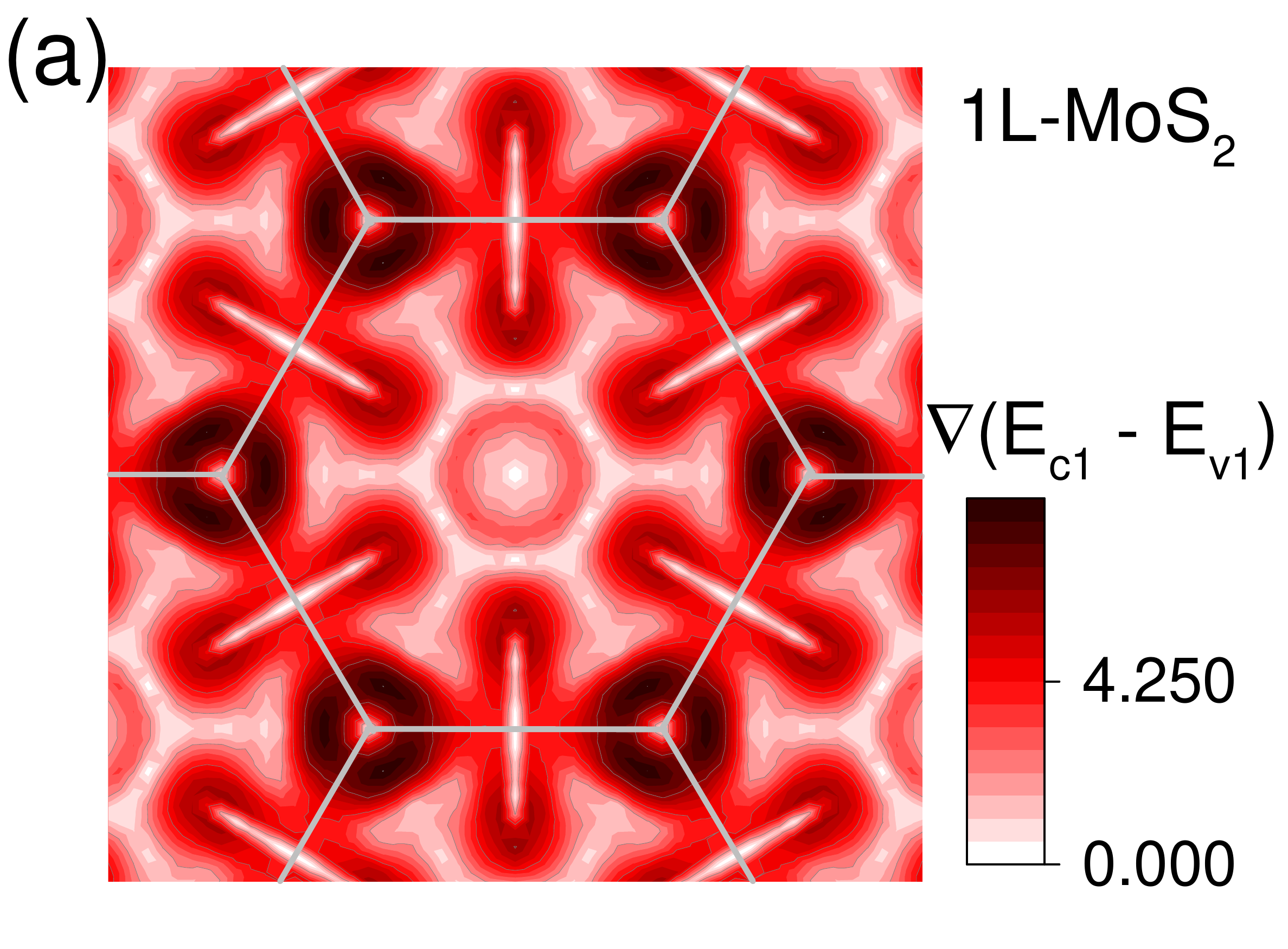}
\includegraphics*[width=0.49\columnwidth]{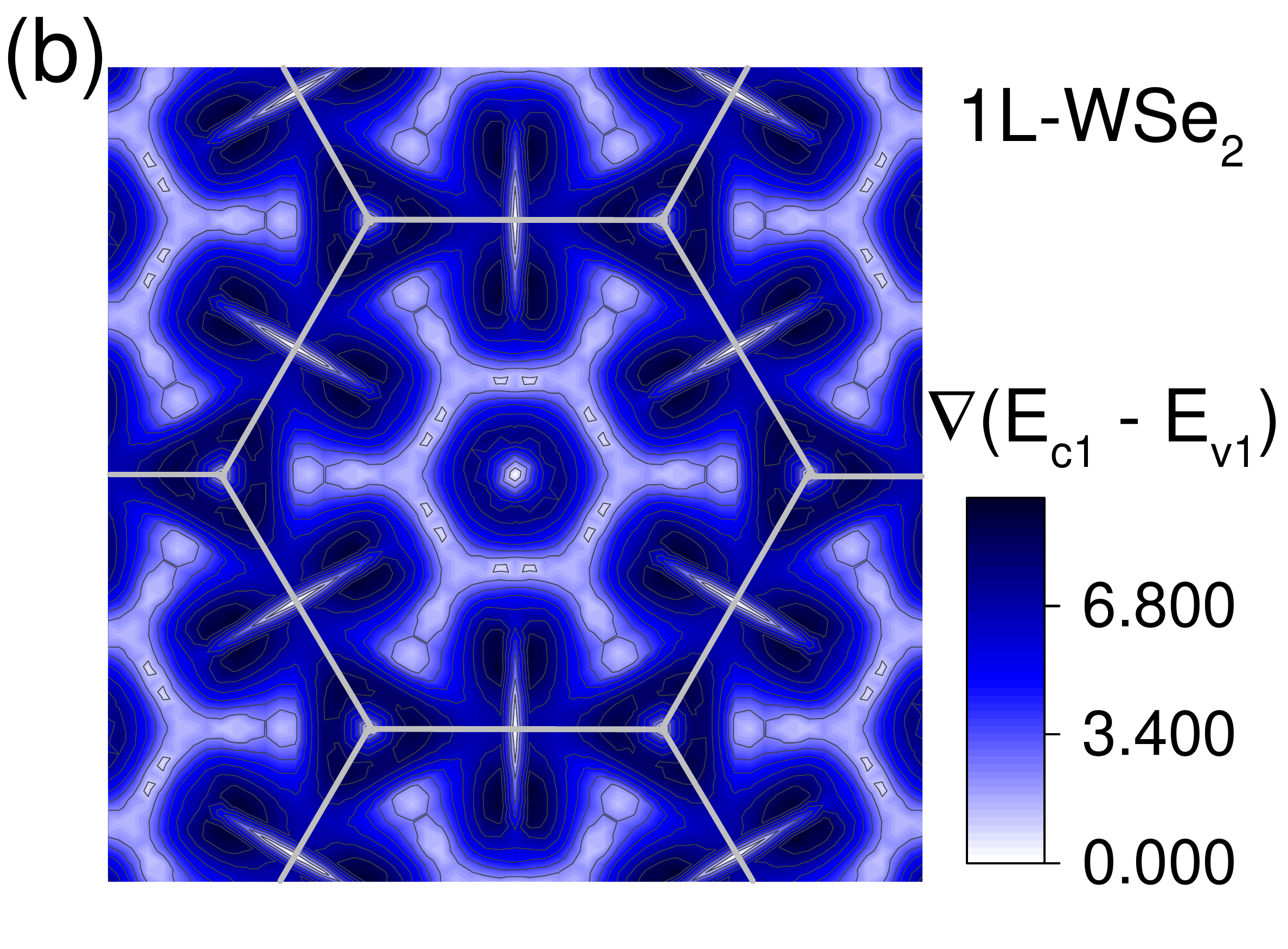}
\caption{\label{fig:nesting} (Color online) Difference between the energy gradients of the highest valence band ($v_1$) and the lowest-energy conduction band ($c_1$) plotted over the Brillouin zone for monolayer WSe$_2$. Similar figures are obtained for MoS$_2$, WS$_2$ and MoSe$_2$ as well. Whiter areas show parts of the Brillouin zone where $v_1$ and $c_1$ are approximately parallel, thus leading to a high joint-density-of-states. }
\end{figure}

For monolayer MoS$_2$ and WS$_2$, the absorption spectra feature an additional prominent peak at an energy of about 2.75\,eV, consistent with the broad '\textsl{C}' feature at this energy in experimentally measured absorption spectra~\cite{MoS2-absorption,WS2-absorption}. Based on a decomposition of the calculated absorption spectra into the contributing transitions, the origin of this feature has been attributed to weakly spin-orbit split transitions between the valence and conduction bands at six points approximately half-way {\blau between the $\Gamma$ point and the 'Q' conduction band valley}~\cite{exciton-paper,qiu-2013,carvalho-band-nesting}, refer to the momentum-resolved exciton wavefunction in Figure~\ref{fig:exciton-wvfns}~(b). {\blau To avoid confusion, we note that it is common in the literature to label this valley by the index $\Lambda$ for the monolayer materials.}

Using the value of the electronic band gap close to these six points, we derived a significant exciton binding energy on the order of 700\,meV for MoS$_2$, which is reflected in a small spatial extend of the exciton wavefunction [right side of Figure~\ref{fig:exciton-wvfns}~(b)]. The location of the \textsl{C} excitations in the Brillouin zone appears somewhat unusual on first glance, as, in contrast to the \textsl{A} and \textsl{B} excitons, they do not correspond to transitions between valence and conduction band extrema. Instead, the \textsl{C} exciton arises from a 'band nesting' of transitions between the approximately parallel valence and conduction bands halfway along the $\Gamma$-$Q$ line, see Figure~\ref{fig:nesting}~(a), thus leading to a high joint-density of states and a correspondingly high optical oscillator strength and optical conductivity~\cite{carvalho-band-nesting}. 

As a consequence of this band nesting condition, it is to be expected that the location in the Brillouin zone, peak energy and optical oscillator strength of the \textsl{C} exciton or corresponding excitations should significantly depend on details of the electronic dispersion. As Figure~\ref{fig:monolayer-bands} shows, the dispersion around the valence band maximum of MoSe$_2$ and WSe$_2$ is flatter and the conduction band dispersion is larger than in the disulfides, suggesting an increased repulsion between the transition metal $d$ and chalcogen $p$ states that make up the valence band ($d_{z^2}$+$p_z$) and the conduction band ($d_{xz,yz}$+$p_{x,y}$) edges at the $\Gamma$ point~\cite{andor-paper}. This weakens the band nesting condition on the $\Gamma$-$Q$ line [refer to Figure~\ref{fig:nesting}~(b) for the example of {\blau monolayer} WSe$_2$] and increases the difference between the transition energy at the band nesting points and the direct band gap of the system. {\blau At the same time}, a set of six new band nesting points appear about half-way on the $Q$-K and $Q$-K' lines. The absorption spectra of MoSe$_2$ and WSe$_2$ feature two particularly bright transitions, which give rise to two prominent broader features above an energy of 2.5\,eV [Figure~\ref{fig:monolayer-abs}]. 
The calculated exciton wavefunctions [Figure~\ref{fig:exciton-wvfns}~(c)] suggest that these excitations have significant contributions from transitions at the 'new' band nesting point and hence can be interpreted as \textsl{C} excitons as well, albeit of a different origin than the \textsl{C} excitons in the disulfides. A significant qualitative consequence arises from the fact that the valence band edge at the band nesting point in monolayer MoSe$_2$ and WSe$_2$ retains some of the spin-orbit splitting from the $K$ and $K'$ points, giving rise to two bright \textsl{C} excitations with a significant energetic splitting, in contrast to the very small splitting in the case of the disulfides. Similar results were found for monolayer MoTe$_2$ as well~\cite{exciton-paper}. 
\begin{figure}
\centering
\includegraphics*[width=0.99\columnwidth]{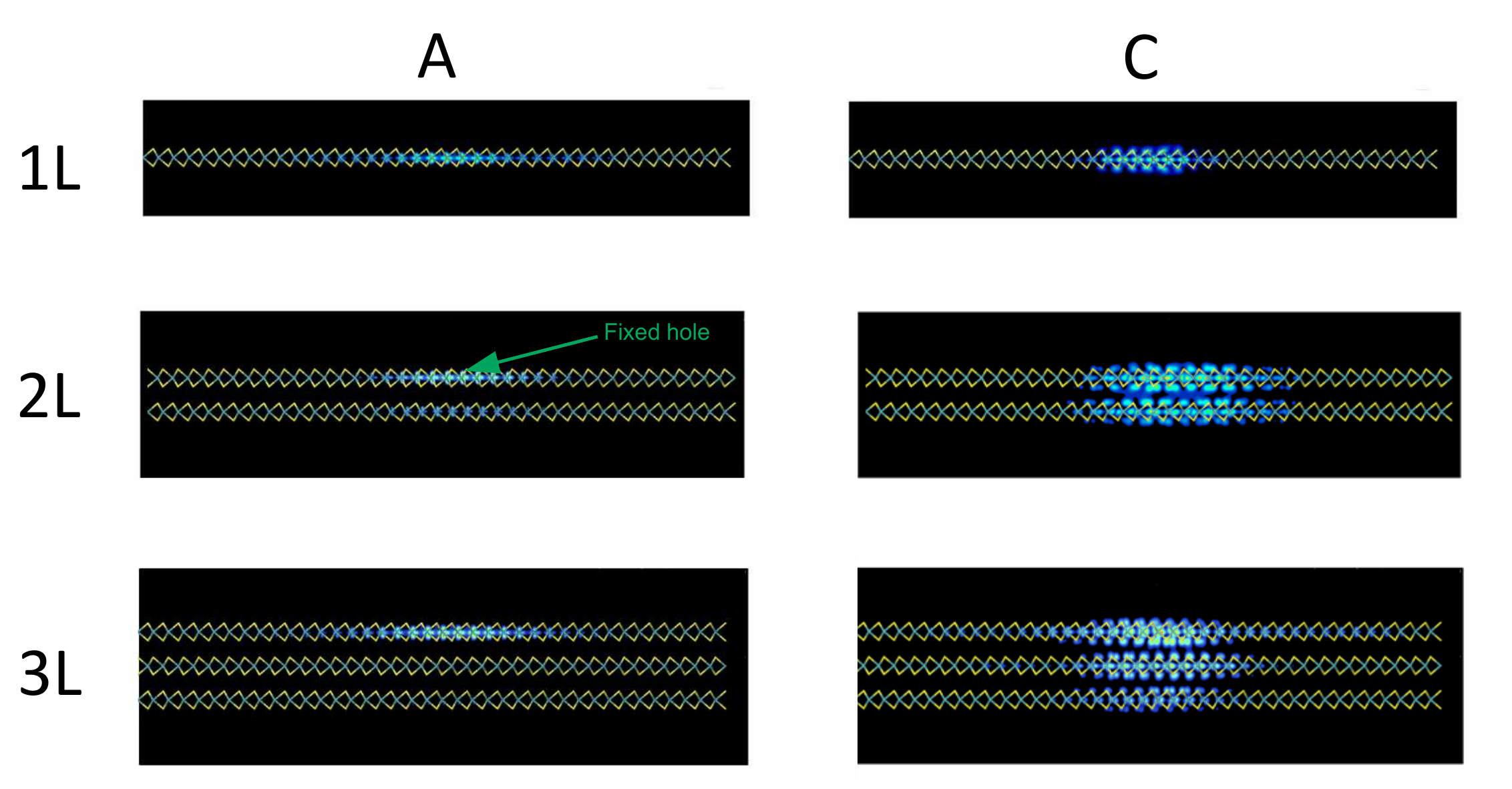}
\caption{\label{fig:AC-excitons} (Color online) Side views of the electron parts of the \textsl{A} and \textsl{C} exciton wavefunctions of mono-, bi-, and trilayer MoS$_2$. In all cases, the hole 
was fixed at a molybdenum atom in the center of the 20x20 supercell used for the plot. Adapted with permission from Ref.~{\blau[\onlinecite{exciton-paper}]}.}
\end{figure}

Another effect on the \textsl{C} excitons of TMDC materials could arise from interlayer or substrate interaction. The significant contributions from chalcogen $p$ states couple at the interface to adjacent materials, for instance a neighboring layer in bilayer Mo$S_2$, and lead to hybridization-induced band splittings. Despite this interlayer hybridization, we reported previously~\cite{exciton-paper} that the resulting changes in the electronic dispersions are sufficiently small to retain the \textsl{C} excitons of the monolayer materials. In this context, the chalcogen $p$ contributions at the band nesting points have interesting consequences in terms of the spatial extent of the \textsl{A} and \textsl{C} exciton wavefunctions. At the K and K' points, the valence band maximum and conduction band minimum are composed almost entirely of transition metal $d_{x^2-y^2}+d_{xy}$ and $d_{z^2}$ orbitals, respectively, causing a negligible overlap and hybridization between the sub{\blau-}bands of each material layer, particularly in light of the additional effect of spin-orbit coupling. As Figure~\ref{fig:AC-excitons} shows for the example of MoS$_2$, this leads to a strong confinement of the electronic part of the exciton wavefunction to the layer where the hole is located. The \textsl{A} exciton in a molybdenum or tungsten TMDC homo-multilayer should hence behave like the \textsl{A} exciton of the corresponding monolayer material. {\blau For the \textsl{C} excitons,} the chalcogen $p$ contributions to the conduction band result in a strong interlayer nature of the excitonic wavefunction, being significantly delocalized over the layers neighboring the layer where the hole is located. This interlayer nature might allow to resonantly couple a TMDC few-layer structure through optical excitation at the energy of \textsl{C} exciton, for instance activating interlayer resonant Raman modes in few-layer TMDCs~\citep{scheuschner-interlayer-modes,staiger-paper,wirtz-quantuminterference} that are not observed for excitation at the \textsl{A} exciton.


\subsection{Electronic bandstructure of MoSe$_2$/WSe$_2$ heterostructures}\label{sec:sec2}
%
%
\begin{figure*}
\centering
\includegraphics*[width=0.9\textwidth]{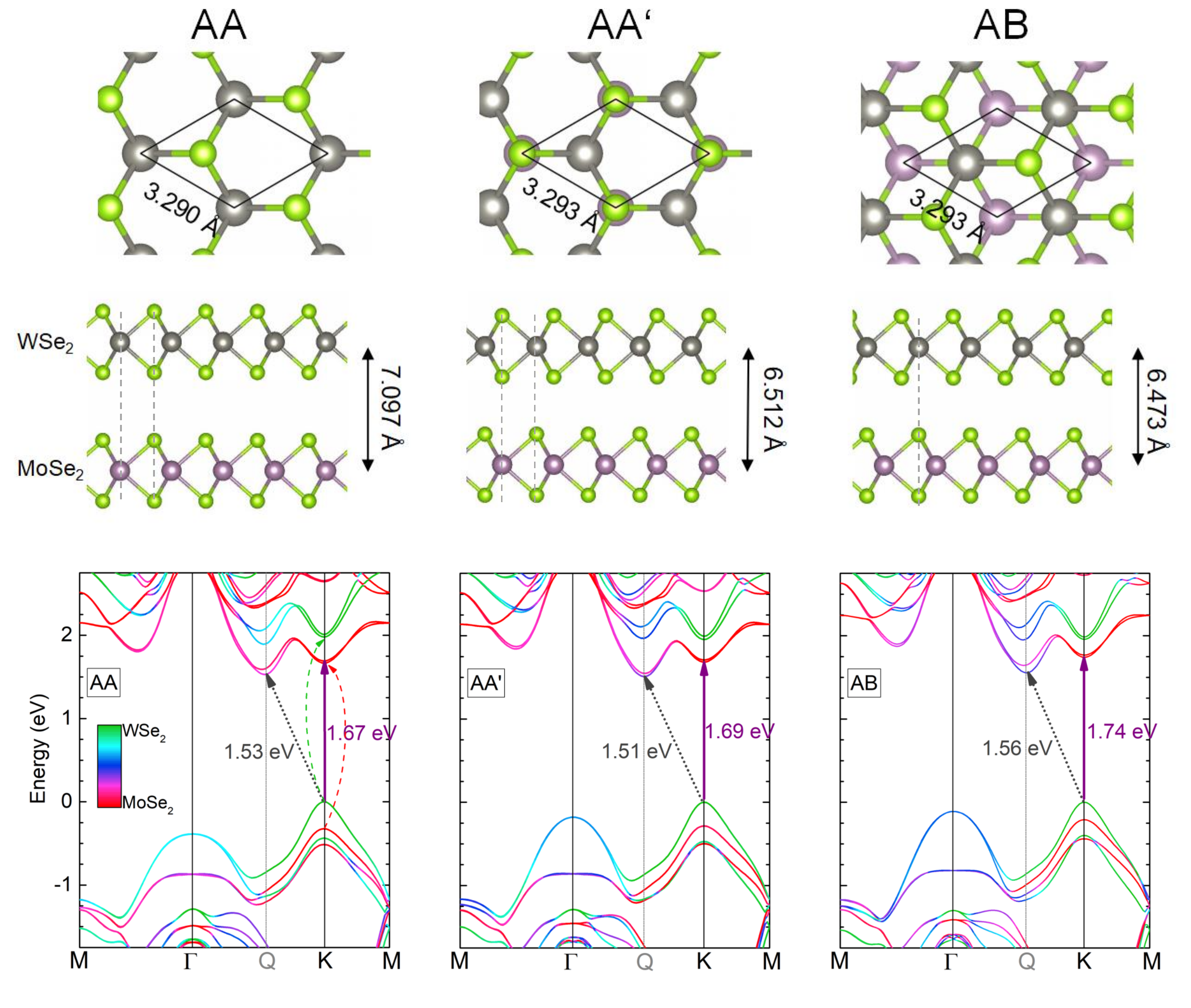}
\caption{\label{fig:geometry_bands} (Color online) Atomic geometries and electronic bandstructures of MoSe$_2$/WSe$_2$ for three different hexagonal stacking orders. The color code for the bandstructures indicates the relative contributions of orbitals of the two sublayers to the bands, and hence the degree of hybridization. Adapted from Ref.~{\blau[\onlinecite{gillen-interlayer}]}. }
\end{figure*}

In the following, we will first summarize our previous results~\cite{gillen-interlayer} on the electronic band structures and excitonic absorption spectra of lattice-commensurate MoSe$_2$/WSe$_2$ heterostructures with different stacking orders. 

Figure~\ref{fig:geometry_bands} shows the electronic bands of vertically stacked MoSe$_2$/WSe$_2$ heterostructures 
obtained from GW calculations, which typically predict the electronic structures of materials with high accuracy, and were reported previously in {\blau Ref.~{\blau[\onlinecite{gillen-interlayer}]}}. To estimate the influence of interlayer coupling on the electronic bandstructures of these heterostructures, we tested three different stacking orders that preserve the hexagonal symmetry. In the AA stacking order, the metal and selenium atoms of the two monolayers are located on top of each others. {\blau AA'} stacking corresponds to the AA stacking order with the upper layer rotated by 60$^\circ$ relative to the lower layer and is the energetically most stable stacking order in homobilayer and bulk TMDCs. {\blau In AB} stacking order, on the other hand, the upper layer is shifted by 1/3($\vec{a}_1$+$\vec{a}_2$) compared {\blau to AA} stacking. Both AA and AB stacking orders are of $R$-type due to their similarity to the stacking order in multilayer 3R-TMDCs, while {\blau AA' stacking} is of $H$-type due to its similarity to the layer arrangement in 2H-TMDCs.

For all three stacking orders, the bands at the K point of the hexagonal Brillouin zone show only a weak hybridization, so that the bands can be directly assigned to the individual layers. This behavior is not unexpected, as the valence and conduction bands at the K and K' points arise largely from transition-metal $d$ states and hence should only weakly {\blau'feel'} the neighboring layer. Indeed, the bands form a type-II alignment, with the (global) valence band of the heterostructure contributed by the WSe$_2$ layer, while the conduction band minima at the K and K' points are located in the MoSe$_2$ layer. The momentum-direct interlayer band gaps between these bands are somewhat stacking-dependent. As it is to be expected that AA and AB are the stacking-orders with the weakest and strongest interlayer coupling for all possible twist angles between the layers, the band gaps between the K and K' of the individual materials should be generally fall within the range of of 1.65-1.75\,eV. Our calculations further indicate that the \text{intra}layer band gaps at the K and K' points of the individual materials are somewhat reduced by about 0.1\,eV due to the dielectric screening provided by the respective other layer.


While the interlayer hybridization is very weak for the band around the K and K' points, the situation is clearly different at other points in the Brillouin zone. The states at the {\blau local} valence band maximum near the $\Gamma$ point and the conduction band minimum near the {\blau Q} point, about half-way along the $\Gamma$-K line, have significant contributions from Se $p$ states and couple strongly between the layers. Similarly to the {\blau homo-bilayer} materials, this interlayer coupling transforms the heterostructure into an indirect semiconductor. The contribution of orbitals from the WSe$_2$ layer to the conduction band minimum is stacking-order dependent and increases from \~20\,\% for AA stacking to about 40\,\% for AB stacking. Our results are in good agreement with other recent studies~\cite{komsa-excitons,wirtz-interlayer} and establish that vertically stacked MoSe$_2$/WSe$_2$ heterobilayers, contrary to previous predictions, only feature a pseudo-type-II band alignment and an indirect fundamental band gap, which has possible implications for the nature and properties of the lowest-energy excitonic states of the material. 
Similar results have been found recently for MoS$_2$/WS$_2$ heterostructures as well~\cite{wirtz-interlayer}. In this case, the indirect fundamental band gap has been predicted to be between the Gamma and the K/K' points, due to the smaller energy separation of the valence band maxima at the $\Gamma$ and K points in the individual monolayer molybdenum and tungsten disulfides.

The electronic bandstructures of the materials suggest two possible origins for the additional 'interlayer' excitonic peaks seen in {\blau PL} experiments: (i) Transitions at the interlayer band gaps at the K and K' points of the Brillouin zone of the heterostructure. The optical strength of these transitions should then strongly depend on the twist angle between the materials, requiring an additional crystal momentum source for twist angles that break hexagonal symmetry. (ii) Transitions over the fundamental band gap, which is indirect for all twist angles. 

\subsection{Momentum-direct interlayer excitons in MoSe$_2$/WSe$_2$ heterostructures}
\begin{figure*}
\centering
\includegraphics*[width=0.9\textwidth]{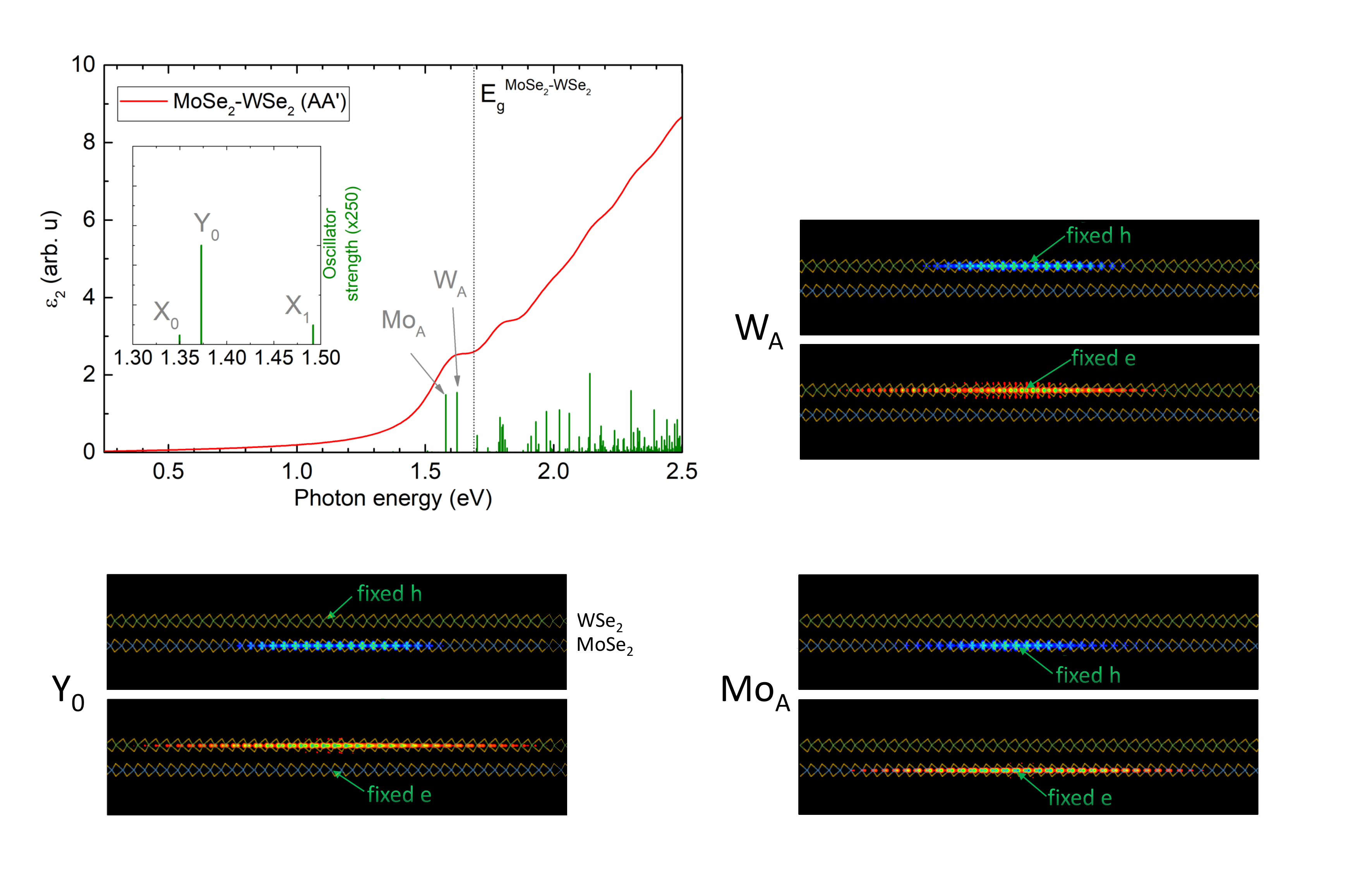}
\caption{\label{fig:AAs_eps_plus_wvfns} (Color online) (a) Simulated absorption spectrum of a MoSe$_2$/WSe$_2$ heterostructure with AA' stacking order. Green lines indicate the individual transitions that give rise to the spectrum. (b) Electron (red density) and hole (blue density) contributions to the wavefunctions of selected excitonic transitions plotted in a supercell of 21x21 unit cells of the heterostructure. The hole (electron) was fixed near a Mo or W site in the middle of the supercell for the plot of the electronic (hole) part. Adapted from Ref.~{\blau[\onlinecite{gillen-interlayer}]}.}
\end{figure*}

We recently reported the simulated absorption spectra of MoSe$_2$/WSe$_2$ heterostructures from solution of the excitonic Bethe-Salpeter equation, which includes effects from excitons with a vanishing center-of-mass momentum~\cite{gillen-interlayer}. Figure~\ref{fig:AAs_eps_plus_wvfns} shows the absorption spectrum for the AA' stacked MoSe$_2$/WSe$_2$ heterostructure with the inclusion of spin-orbit coupling effects. The onset of absorbance is dominated by two transitions, $Mo_A$ and $W_A$, which can readily {\blau be} attributed to the $A$ excitons of individual monolayer MoSe$_2$ and WSe$_2$: our analysis of the {\blau k-resolved contributions to the exciton wavefunction} shows that $Mo_A$ and $W_A$ exclusively arise from intralayer transitions between the MoSe$_2$ (WSe$_2$) bands at the K and K' points. Further, the electron and hole parts of the excitonic wavefunctions show that the corresponding electron-hole pairs are completely confined to the MoSe$_2$ and WSe$_2$ layers, respectively [Figure~\ref{fig:AAs_eps_plus_wvfns}]. In principle, one could directly derive the exciton binding energies from a comparison of the peak positions in the calculated absorption spectrum with the intralayer band gaps from Figure~\ref{fig:geometry_bands}. However, while the 21x21 k-point grid used in the calculation is sufficient to give a very accurate qualitative picture of the excitonic spectrum, it does not yield fully converged peak positions and exciton binding energies. A second calculation using a denser 33x33 k-point grid (without inclusion of spin-orbit coupling) yields binding energies of about 310\,meV and 290\,meV for the $Mo_A$ and $W_A$ transitions, respectively, for the AA' stacking order. This suggests that the additional dielectric screening induced by the neighboring layer reduced the binding energies of the intralayer $A$ excitons by about 150\,meV.
%
\begin{table*}
\caption{\label{tab:tab_lifetimes} Averaged radiative lifetimes of inter- and intralayer excitons in MoSe$_2$/WSe$_2$ heterostructures with different stacking orders for a temperature of 4\,K ($\left\langle\tau\right\rangle$) calculated using Equations~\ref{eq:taupar} and~\ref{eq:tauperp} and the excitonic energies and wavefunctions from our BSE simulations. ILX indicate the lowest-energy 'bright' interlayer exciton ($X_0$) for AA and AB stacking; for AA' stacking, lifetimes for both the dark $X_0$ and the bright $Y_0$ are given. All interlayer exciton lifetimes correspond to emission with light polarization parallel to the plane ($\left\langle\tau_s^\parallel\right\rangle$), except for AB stacking, where emission of out-of-plane polarized light is predominant ($\left\langle\tau_s^\perp\right\rangle$).}
\begin{ruledtabular}
\begin{tabular}{ c  c  c  c  c }
\multirow{2}{*}{Exciton} & \multirow{2}{*}{Lifetimes} & & Stacking order & \\
 &  & AA & AA' & AB\\
\hline
\multirow{4}{*}{ILX}    & \multirow{2}{*}{$\tilde{\tau}_s$(0)}                                      & \multirow{2}{*}{65.1\,ps}  & 0.34\,ns ($X_0$) & \multirow{2}{*}{737.6\,ps {\blau ($X_0$)}}\\
                        &                                                                           &                            & 31.6\,ps ($Y_0$) & \\
                        & {\blau \multirow{2}{*}{$\left\langle\tau_s^{\parallel}\right\rangle$}}    & \multirow{2}{*}{10.5\,ns}  & 53.8\,ns ($X_0$) & \multirow{2}{*}{}\\
                        &                                                                           &                            & 5.34\,ns ($Y_0$) & \\												
                        & {\blau $\left\langle\tau_s^{\perp}\right\rangle$}                         &                            &         & {\blau 213.2\,ns ($X_0$)}\\
												\hline
\multirow{2}{*}{$Mo_A$} & $\tilde{\tau}_s$(0)                                                       & 0.34\,ps & 0.37\,ps & 0.41\,ps\\
                        & $\left\langle\tau_s^{\parallel}\right\rangle$                               & 46.3\,ps & 50.6\,ps & 57.2\,ps\\
\hline
\multirow{2}{*}{$W_A$}  & $\tilde{\tau}_s$(0)                                                       & 0.30\,ps & 0.34\,ps & 0.35\,ps\\
                        & $\left\langle\tau_s^{\parallel}\right\rangle$                               & 25.1\,ps & 29.5\,ps & 29.0\,ps\\
\end{tabular}
\end{ruledtabular}
\end{table*}

In addition to these intralayer excitons, the computed absorption spectra also reveal two series of peaks of small optical oscillator strength at lower energies, see inset of Figure~\ref{fig:AAs_eps_plus_wvfns}. These peaks arise from interlayer transitions between the WSe$_2$-dominated global valence band maximum of the heterostructure to the spin-orbit-split conduction band minimum at the K and K' points; the $X_n$ and $Y_n$ peaks correspond to transitions to the lower energy and to the higher energy band, respectively. In the AA' stacking order, the monolayer Brillouin zones of MoSe$_2$ and WSe$_2$ are rotated by 60$^\circ$ relative to each other, i.e. the K points of one material coincide with the K' points of the other material. As a consequence, the $Y_0$ transition should be brighter than the lower energy $X_0$ transition due to spin-conservation during the optical transition, while the $X_0$ transition should be dark in heterostructures with H-type stacking order. The excitonic wavefunction of the $Y_0$ transition is shown in Figure~\ref{fig:AAs_eps_plus_wvfns}. In contrast to the $Mo_A$ and $W_A$ transitions, the exciton wavefunction shows a clear spatial separation of the electron and hole pairs between the layers. From calculations with a denser k-point grid, we derive a rather large exciton binding energy of about 250\,meV, which is of similar magnitude as that of the intralayer excitons. Our calculations thus suggest that the $X_n$ and $Y_n$ peaks form two Rydberg series of interlayer excitons. Similar results were found for AA and AB stacking orders as well~\cite{gillen-interlayer}. 
Our calculations hence support the initial assignment of these PL peaks to interlayer excitons from momentum-direct excitations over the interlayer band gap at the K and K' points of the heterostructure. The low oscillator strength of these transitions is in agreement with recent photocurrent measurements in MoSe$_2$/WSe$_2$ p-n junctions, which suggested that the oscillator strength of the intralayer excitons is 200 times larger than that of the interlayer excitons~\cite{ross-MoSe2WSe2-pn}. On the other hand, Torun \emph{et al}. recently pointed out that the high exciton lifetime due to the spatial separation of the electron-hole pair, in complete agreement with experimental investigation of the interlayer peak~\cite{wirtz-interlayer}, is much larger than the exciton thermalization time. The high interlayer peak intensity in PL spectra can hence be understood from a Boltzmann-type thermal occupation of excitonic states, which compensates for the low oscillator strength of the interlayer excitons due to the energy separation to the intralayer excitonic states. 

Due to depolarization effects in out-of-plane direction, only light with parallel polarization to the materials surface should appreciably couple to excitons in 2D materials~\cite{palumno-2015}. This is indeed the case for the AA and AA' stacking orders, where either the $X_0$ or the $Y_0$ transitions are 'bright' for parallel light polarization, but dark for 'perpendicular' light polarization that is parallel to the surface normal vector (see supplementary material of Ref.~{\blau[\onlinecite{gillen-interlayer}]} for a comparison). On the other hand, neither $X_0$ nor $Y_0$ show any appreciable oscillator strength for AB stacking~\cite{gillen-interlayer}. A detailed analysis revealed that this behavior can be traced back to variations in the optical selection rules for AB stacking compared to AA or AA' stacking. The oscillator strength of the interlayer excitons largely arises from transitions between small contributions of MoSe$_2$ $d$ states mixed into the valence band maximum at the K point into the MoSe$_2$ dominated conduction band. For AB stacking these small contributions come from Mo $d_z$ states, which causes the interlayer transitions to be optically active for perpendicular polarized light, albeit with a much smaller optical oscillator strength than for AA or AA stacking. These results suggest that the stacking order can have a marked effect on the coupling of interlayer excitons to polarized light. As we show in the following, the small oscillator strength also has strong implicatons for the radiative lifetime of interlayer excitons for AB stacked MoSe$_2$/WSe$_2$ heterostructures.

Starting from Fermi's Golden rule, {\blau an expression for the computation of radiative lifetimes of excitonic states in 2D materials from the excitonic oscillator strengths can be derived~\cite{urbaszek-2014,palumno-2015}. Following this approach, the radiative recombination rate of the excitonic state $s$ with a small center-of-mass momentum $\mathbf{Q}$ emitting light polarized parallel to the 2D material plane is given by}

\begin{equation*}
\gamma_s^{\parallel}(\mathbf{Q}) = \gamma_s\cdot\left( \sqrt{ 1-\left( \frac{\hbar c^2Q}{E_s(\mathbf{Q})}\right)^2} + \frac{1}{2}\frac{\left(\frac{\hbar c \left(Q_x-Q_y\right)}{E_s(\mathbf{Q})}\right)^2}{\sqrt{ 1-\left( \frac{\hbar c^2Q}{E_s(\mathbf{Q})}\right)^2}} \right),
\end{equation*}

with a zero-momentum lifetime

\begin{equation}
\tilde{\tau}_s(0)=\gamma_s^{-1}=\frac{\epsilon_0A_{uc}\hbar^2c}{e^2E_s(0)\mu_{s,\parallel}^2}
\end{equation}

Thermal averaging of the radiative transition rate $\gamma_s(G)$ and inversion gives the averaged lifetime of exciton state $s$ for temperature $T$~\cite{palumno-2015},

\begin{equation}
\left\langle\tau_s^{\parallel}\right\rangle\approx\frac{3}{4}\tilde{\tau}_s\frac{2M_sc^2}{E_s^2(0)}k_BT.\label{eq:taupar}
\end{equation}

These {\blau expressions} assume a quadratic exciton dispersion, $E_s(Q)=E_s(0)+\frac{\hbar^2Q^2}{2M_s}$. The necessary ingredients, the zero-momentum exciton binding energies $E_s(0)$, the effective exciton mass $M_s=m_e^{*,s}+m_h^{*,s}$ and the transition dipole 
$\mu_{s,\parallel}^2=\frac{\hbar^2}{N_k}\left|\left\langle\psi_0|\mathbf{r}^\parallel|\psi_s\right\rangle\right|^2$ can be directly obtained from the GW bandstructures and solution of the BSE. $N_k$ is the total number of k-points in the grid used for the solution of the BSE.
For AB stacking, Equation~\ref{eq:taupar} is not applicable, as the recombination of interlayer excitons should not couple to parallel polarized light. We hence followed the approach used by Palumno \emph{et al.}~\cite{palumno-2015} to arrive at the following expressions of the momentum-dependent recombination rate and the thermally averaged lifetime for perpendicular polarized light:
\begin{align}
\gamma_s^{\perp}(Q) &= \frac{E_s\left(0\right)}{2\hbar c}\tilde{\tau}_s^{-1}(0)\cdot\frac{Q^2}{Q_0^3\sqrt{1-\frac{Q^2}{Q_0^2}}}\nonumber\\
\left\langle\tau_s^\perp\right\rangle &\approx\frac{3}{2}\tilde{\tau}_s(0)\frac{2M_sc^2}{E_s^2(0)}k_BT.\label{eq:tauperp}
\end{align}
Table~\ref{tab:tab_lifetimes} shows the calculated lifetimes for MoSe$_2$/WSe$_2$ heterostructures with three different stacking orders. The effective masses of the valence and conduction bands at $K$ are only weakly dependent on the stacking order and do not significantly affect the predicted lifetimes. For the intralayer excitons, the calculated optical oscillator strength decreases along the sequence AA'$\rightarrow$AA$\rightarrow$AB, which, according to Equations~\ref{eq:taupar} and~\ref{eq:tauperp}, causes a corresponding increase of the predicted radiative lifetimes. As expected from experimental data~\cite{WSe2MoSe2-2,miller-MoSe2WSe2,lifetime-1,lifetime-2,Choi-2018}, the lifetimes of the interlayer excitons are predicted to be in the nanosecond range even for very low temperatures and thus significantly larger than those of the intralayer excitons. The stacking order has a marked effect on the predicted radiative lifetimes{\blau. For AA' stacking, our calculated zero-momentum lifetime $\tilde{\tau}_s$(0) is aproximately half the value reported recently from simulations using a combined Dirac-Bloch and gap equations approach~\cite{meckbach} and substantially larger (by a factor 3) than those reported for a bilayer MoSe$_2$/WSe$_2$ heterostructure by Palumno \emph{et al}~\cite{palumno-2015}. In the AA stacked heterostructure, the lifetime of the bright $X_0$ transition  is more than twice as long as that of the $Y_0$ transition for AA' stacking.} These differences might be related to different absolute magnitudes of the optical oscillator strengths. 
Due to the low oscillator strength even for perpendicular polarized light, we predict the the zero-momentum interlayer excitons for AB stacking to be extremely long-lived in terms of radiative decay time, on the order of microseconds; correspondingly, we expect non-radiative decay channels to be prevalent for interlayer excitons in AB stacked MoSe$_2$/WSe$_2$ heterostructures.

\subsection{Momentum-indirect interlayer excitons in TMDC heterostructures}
We will now turn towards the possible contribution of excitons with \emph{\blau non-vanishing} center-of-mass momentum, for example transitions related to the fundamental indirect band gap of TMDC heterostructures, to the {\blau 'interlayer'} peaks observed in PL experiments. Such an indirect transition has been proposed recently to be the origin of the prominent peak at 1.6\,eV observed for lattice-incommensurate Mo$S_2$/WSe$_2$ heterostructures, in this case between the valence band maximum at {\blau the }$\Gamma$ point of the heterostructure and the local minimum related to the K point of the Mo$S_2$ sublayer~\cite{kunstmann-interlayer-excitons}. An additional feature, which we attributed to the (momentum-indirect) $K^{\mbox{\tiny WSe$_2$}}\rightarrow K^{\mbox{\tiny MoS$_2$}}$ transition, was observed at an energy of about 1\,eV, in excellent agreement with theoretical simulations of the expected peak position~\cite{Karni-Mos2Wse2}.
\begin{table*}
\caption{\label{tab:tab_indirect_ILX} Calculated binding energies and predicted peak positions of various momentum-direct and momentum-indirect excitons in AA'-stacked MoSe$_2$/WSe$_2$ heterobilayers from use of a simple tight-binding model~\cite{kunstmann-interlayer-excitons} and explicit solution of the Bethe-Salpeter Equation. {\blau For the intralayer and momentum-direct interlayer excitons ($K^v$-$K^c$), the values correspond to the lowest-energy optically active transitions; otherwise the lowest-energy exciton for a particular momentum-transfer is listed. 
The corresponding electronic band gaps are the transition energies without inclusion of electron-hole interaction effects and can be directly derived from the calculated electronic bandstructures plotted in {\blau Figure}~\ref{fig:geometry_bands}.} All values are given in units of eV.}
\begin{ruledtabular}
\begin{tabular}{ c  c  c  c  c  c  c  c }
& Intralayer MoSe$_2$ & Intralayer WSe$_2$ & $K^v$-$K^c$ & $K^v$-$K'^c$ & $K^v$-$Q^c$ & $\Gamma^v$-$Q^c$ & $\Gamma^v$-$K^c$\\
\hline
{\blau Electronic} & \multirow{2}{*}{1.98} & \multirow{2}{*}{1.97} & \multirow{2}{*}{1.71} & \multirow{2}{*}{1.69} & \multirow{2}{*}{1.51} & \multirow{2}{*}{1.69} & \multirow{2}{*}{1.87}\\
band gap &  &  &  &  &  &  & \\
\hline
{\blau Exciton binding} & \multirow{2}{*}{0.41} & \multirow{2}{*}{0.36} & \multirow{2}{*}{0.24} & \multirow{2}{*}{0.24} & \multirow{2}{*}{0.35} & \multirow{2}{*}{0.52} & \multirow{2}{*}{0.62}\\
{\blau energy (model)}  &      &      &      &      &      &      &     \\
\hline
{\blau Exciton binding} & \multirow{2}{*}{0.31\footnotemark[1]\footnotetext{33x33x1 k-point grid without spin-orbit coupling}} & \multirow{2}{*}{0.28\footnotemark[1]} & \multirow{2}{*}{0.25\footnotemark[1]} & \multirow{2}{*}{0.23\footnotemark[2]\footnotetext{36x36x1 k-point grid without spin-orbit coupling}} & \multirow{2}{*}{0.22\footnotemark[2]} & \multirow{2}{*}{0.29\footnotemark[2]} & \multirow{2}{*}{0.25\footnotemark[2]}\\
{\blau energy (BSE)}  &      &      &      &      &      &      &     \\
\hline
Peak {\blau position model} & 1.57 & 1.61 & 1.48 & 1.45 & 1.16 & 1.17 & 1.25\\
              {\blau BSE  } & 1.67 & 1.7  & 1.47 & 1.46 & 1.29 & 1.4  & 1.62\\
\end{tabular}
\end{ruledtabular}
\end{table*}

A simple and {\blau computationally} efficient way to estimate the binding energies of excitons with non-zero center-of-mass momentum was proposed recently by Kunstmann \emph{et al.}, who used a four-band tight binding model together with input from DFT calculations to compare the predicted exciton peak positions in MoS$_2$/WSe$_2$ heterostructures with PL measurements~\cite{kunstmann-interlayer-excitons}. 
The resulting estimated binding energies of the 'momentum-direct' $K\rightarrow K$ exciton and the three 'momentum-indirect' $K\rightarrow Q$, $\Gamma\rightarrow Q$, and $\Gamma\rightarrow K$ excitons, as well as the predicted peak positions, are given in Table~\ref{tab:tab_indirect_ILX} for a MoSe$_2$/WSe$_2$ heterostructure with AA' stacking order {\blau [refer to Sec.~2 of the SI for the input parameters to the tight-binding model extracted from our GW calculations]}. The model appears to give an estimate of the exciton binding energy of the $K\rightarrow K$ exciton that is in very good agreement with the BSE results, albeit at much lower computational cost. On the other hand, the model suggests that all three considered 'momentum-indirect' excitons have binding energies similar to or larger than those of the intralayer excitons, which is expected to shift the $K\rightarrow Q$ {\blau and} $\Gamma\rightarrow Q$ peaks below the experimentally observed peak positions. The predicted peak position of the $\Gamma\rightarrow K$ transition is in somewhat better agreement. 
However, the different valence and conduction {\blau band} extrema enter the model purely through their effective masses, band offsets and interlayer coupling parameters. Further, the dielectric screening entering the model is purely local and thus contains momentum-dependent screening effects only in an averaged way. 

The Bethe-Salpeter Equation {\blau should be} a more accurate and consistent approach, which also {\blau explicitly takes into account non-local interaction effects}. However, the typical implementations of the BSE in solid state codes are aimed at the simulation of optical absorption spectra and hence only include contributions with vanishing center-of-mass momentum ($\mathbf{Q}=\mathbf{0}^+$). We hence extended the widely used YAMBO code~\cite{yambo} to solve the more general Bethe-Salpeter Equation~\cite{bgw-2}
\begin{align}
&\left(\epsilon_{c,\mathbf{k}+\mathbf{Q}}-\epsilon_{v,\mathbf{k}}\right)A_{vc\mathbf{k}\mathbf{Q}}^S+\sum_{v'c'\mathbf{k}'}\left\langle vc\mathbf{k}+\mathbf{Q}\left|K^d+K^x\right|v'c'\mathbf{k}'\right\rangle\nonumber\\
&\quad =E^S_QA_{vc\mathbf{k}\mathbf{Q}}^S\label{eq:3}
\end{align}
for a finite exciton center-of-mass momentum $\mathbf{Q}$. The band energies $\epsilon_n\mathbf{k}$ are typically quasiparticle energies obtained from G$_0$W$_0$ calculations. The exchange kernel 
\begin{align}
\left\langle vc\mathbf{k}+\mathbf{Q}\left|K^x\right|v'c'\mathbf{k}'\right\rangle &= \sum_{\mathbf{G},\mathbf{G}'}\rho_{cv}\left(\mathbf{k}+\mathbf{Q},\mathbf{Q},\mathbf{G}\right)v_{\mathbf{Q}-\mathbf{G}}\nonumber\\
&\quad \times\rho_{cv{'}}^*\left(\mathbf{k}'+\mathbf{Q},\mathbf{Q},\mathbf{G'}\right),\nonumber
\end{align}
couples spin-conserving transitions, while the {\blau 'direct'} kernel 
\begin{align}
\left\langle vc\mathbf{k}+\mathbf{Q}\left|K^d\right|v'c'\mathbf{k}'\right\rangle &= \sum_{\mathbf{G},\mathbf{G}'}\rho_{cc'}\left(\mathbf{k}+\mathbf{Q},\mathbf{q},\mathbf{G}\right)W_{\mathbf{GG}'}(\mathbf{q})\nonumber\\
&\quad \times\rho_{vv{'}}^*\left(\mathbf{k},\mathbf{q},\mathbf{G'}\right)\nonumber
\end{align}
determines the formation of bound electron-hole pairs. Here, $v$ is the bare Coulomb interaction, $W_{\mathbf{G}\mathbf{G}'}(\mathbf{q})=\epsilon^{-1}_{\mathbf{G}\mathbf{G}'}(\mathbf{q})\sqrt{v(\mathbf{q}+\mathbf{G})}\sqrt{v(\mathbf{q}+\mathbf{G}')}$ is the screened Coulomb interaction, $\rho_{nn'}\left(\mathbf{k},\mathbf{q},G\right)=\left\langle n\mathbf{k}\left|e^{-i\mathbf{G}\cdot\mathbf{r}}\right|n'\mathbf{k}-\mathbf{q}\right\rangle$ are form factors of the underlying interband transitions, and $\mathbf{q}=\mathbf{k}-\mathbf{k}'$ {\blau is a transferred momentum}. 
The typical implementation of the BSE in most solid state physics codes is recovered by setting $\mathbf{Q}$=$\mathbf{0}$.

In order to ensure meaningful quantitative results, we solved equation~\ref{eq:3} using a grid of 36x36x1 k-points, neglecting effects from spin-orbit interaction. The obtained exciton binding energies are reported in Table~\ref{tab:tab_indirect_ILX}. In stark contrast to the model calculations, the BSE predicts the binding energies of the excitons related to indirect electronic transitions in the MoSe$_2$/WSe$_2$ heterostructure to have very similar magnitudes as those of the $K\rightarrow K$ exciton, in the range of 0.2-0.3\,eV. We now find that a possible peak from the $\Gamma\rightarrow K$ transition should be expected close to the intralayer excitonic peaks, suggesting that $\Gamma\rightarrow K$ indeed should not contribute to the interlayer excitonic peak at lower energies. On the other hand, the peak energies of both the $\Gamma\rightarrow Q$ and $K\rightarrow Q$ excitons are predicted to be quite close to those of the $K\rightarrow K$ exciton. As both excitons involve the $Q$-point conduction band minimum, inclusion of spin-orbit interaction should give rise to two peaks {\blau that are} split by {\blau an energy of about} 35\,meV in both cases{\blau, in agreement with the recently reported experimental observation of momentum-indirect interlayer excitons~\cite{hanbicki-MoSe2WSe2}.}

\begin{figure}
\centering
\includegraphics*[width=0.9\columnwidth]{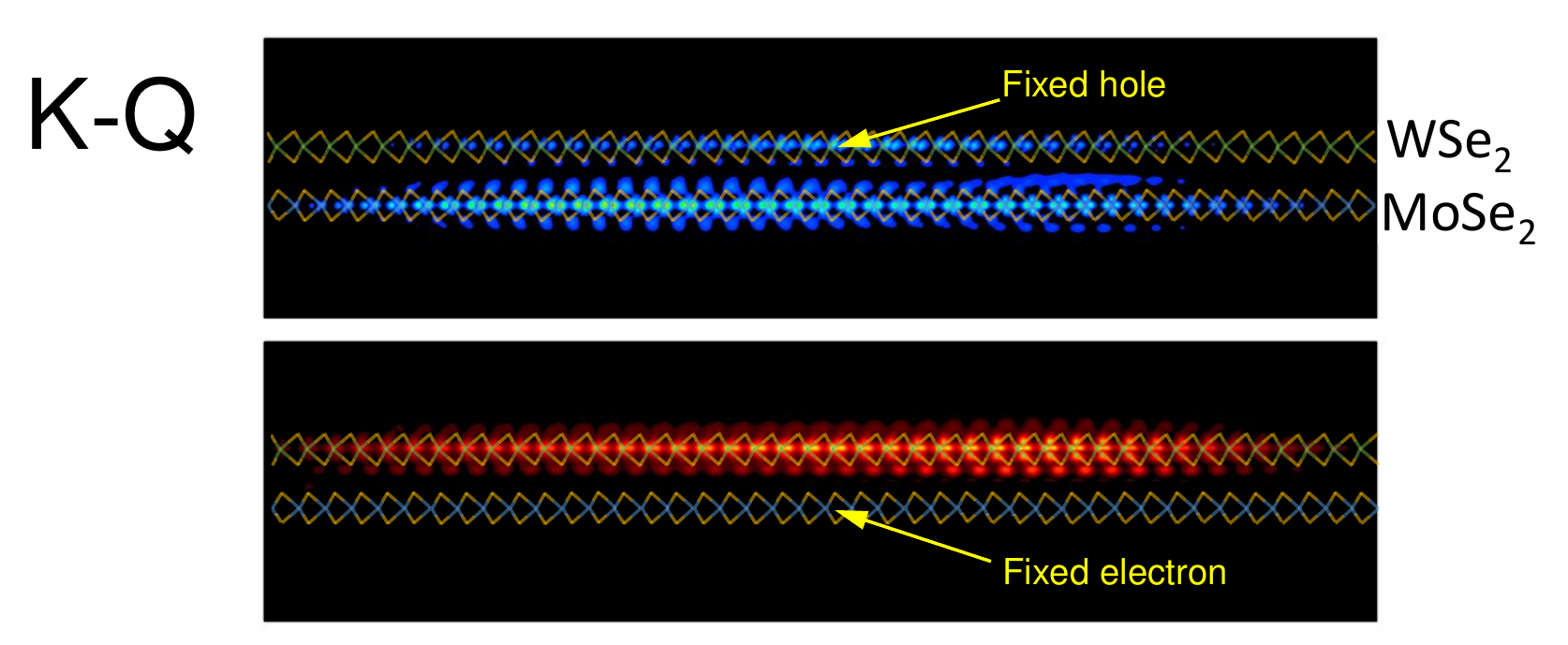}
\includegraphics*[width=0.9\columnwidth]{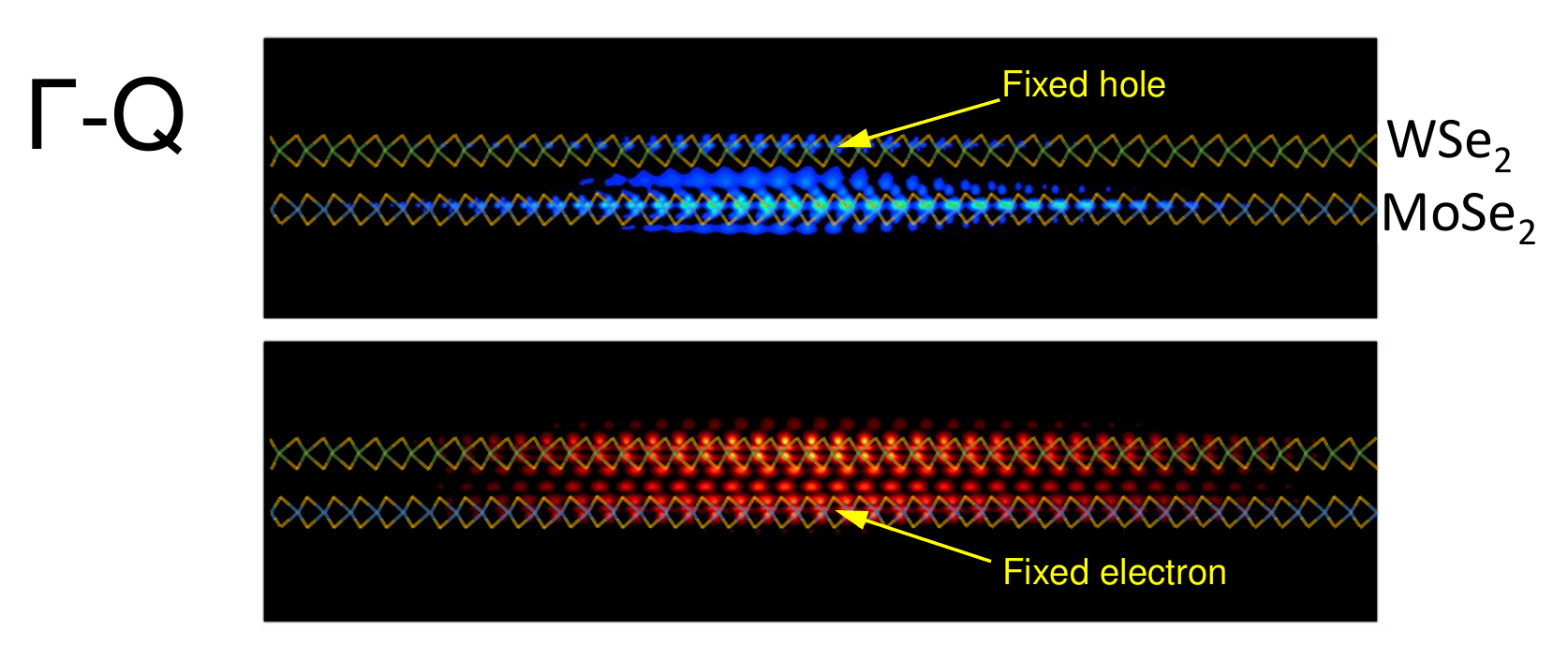}
\includegraphics*[width=0.9\columnwidth]{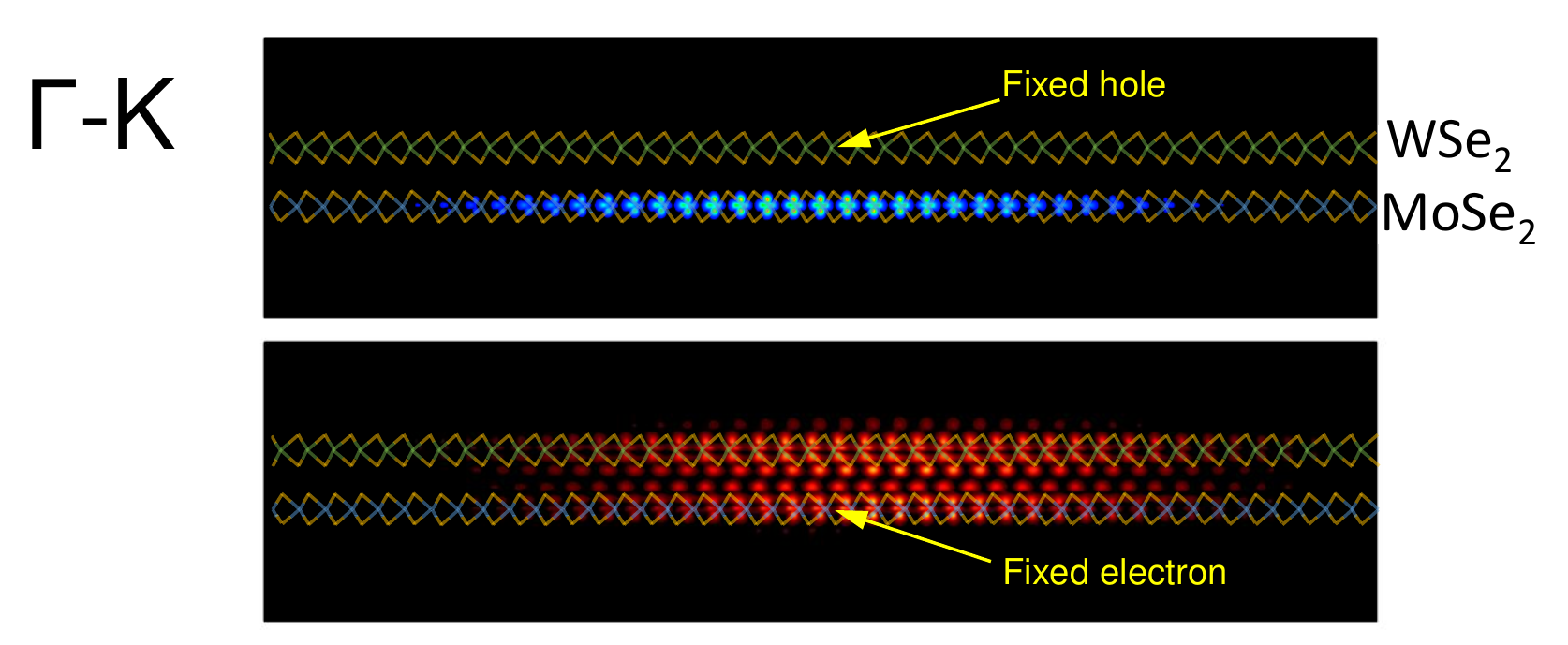}
\caption{\label{fig:MoSe2-WSe2_indir_wvfns} (Color online) Excitonic wavefunctions of various momentum-indirect excitons in AA'-stacked MoSe$_2$/WSe$_2$ heterostructures, divided into electron (red density) and hole (blue density) contributions.}
\end{figure}

An advantage of the BSE is the access to the exciton wavefunctions for each predicted excitonic state in the system, in addition to the peak positions and exciton binding energies. Figure~\ref{fig:MoSe2-WSe2_indir_wvfns} shows the calculated electron and hole parts of the excitonic wavefunctions of the three considered 'indirect' excitons. 
Based on the electronic bandstructures shown in Figure~\ref{fig:geometry_bands}, one would expect that interlayer excitons have a much weaker layer confinement, and thus spatial separation, of the bound electrons and holes if $\Gamma$ or $Q$ points are involved, due to the significant interlayer hybridization effects at these points in the Brillouin zone. {\blau Our calculations suggest that this assumption is indeed true for the hole part of the exciton wavefunction: For a hole located at the K point of the Brillouin zone, the corresponding wavefunction is confined to the WSe$_2$ layer, while for a hole located at the $\Gamma$ point, the corresponding wavefunction has a strong interlayer nature with approximately equal weight on the two layers. This suggests that excitons involving $\Gamma$-point holes should exhibit characteristically small exciton dipoles compared to the 'momentum-direct' $K\rightarrow K$ or $K\rightarrow K'$ excitons that exhibit a distinct interlayer charge separation. For excitons involving the $Q$ point conduction band valley, the electron part of the exciton wavefunctions exhibits a small spilling of the electron into the WSe$_2$ layer, while the larger part of the wavefunction is located in the MoSe$_2$ layer. Based on an integration of the in-plane averaged electron part of the exciton wavefunction along the out-of-plane direction, we estimate this spill-over to be about 30\,\% of the electron wavefunction if spin-orbit interaction (SOI) is neglected. This coincides well with the relative contribution of electronic orbitals in the WSe$_2$ layers to the Q conduction band valley (29\,\%, without SOI), which we estimated using the same method for the Kohn-Sham orbital of the conduction band minimum obtained from DFT. In general, the degree of charge separation for momentum-indirect interlayer excitons should depend significantly on the local stacking order: For an AA stacking order, which we expect to constitute the lowest limit in terms of interlayer hybridization, the contribution of the WSe$_2$ layer to the DFT wavefunction of the Q valley conduction band minimum is reduced to 16\,\% (without SOI). We note that inclusion of SOI further increases the WSe$_2$ layer weight to the DFT orbitals to 29\,\% for AA stacking and 40\,\% for AA' stacking, respectively, in good agreement with the composition shown in {\blau Figure}~\ref{fig:geometry_bands}.}

\begin{figure}
\centering
\includegraphics*[width=0.99\columnwidth]{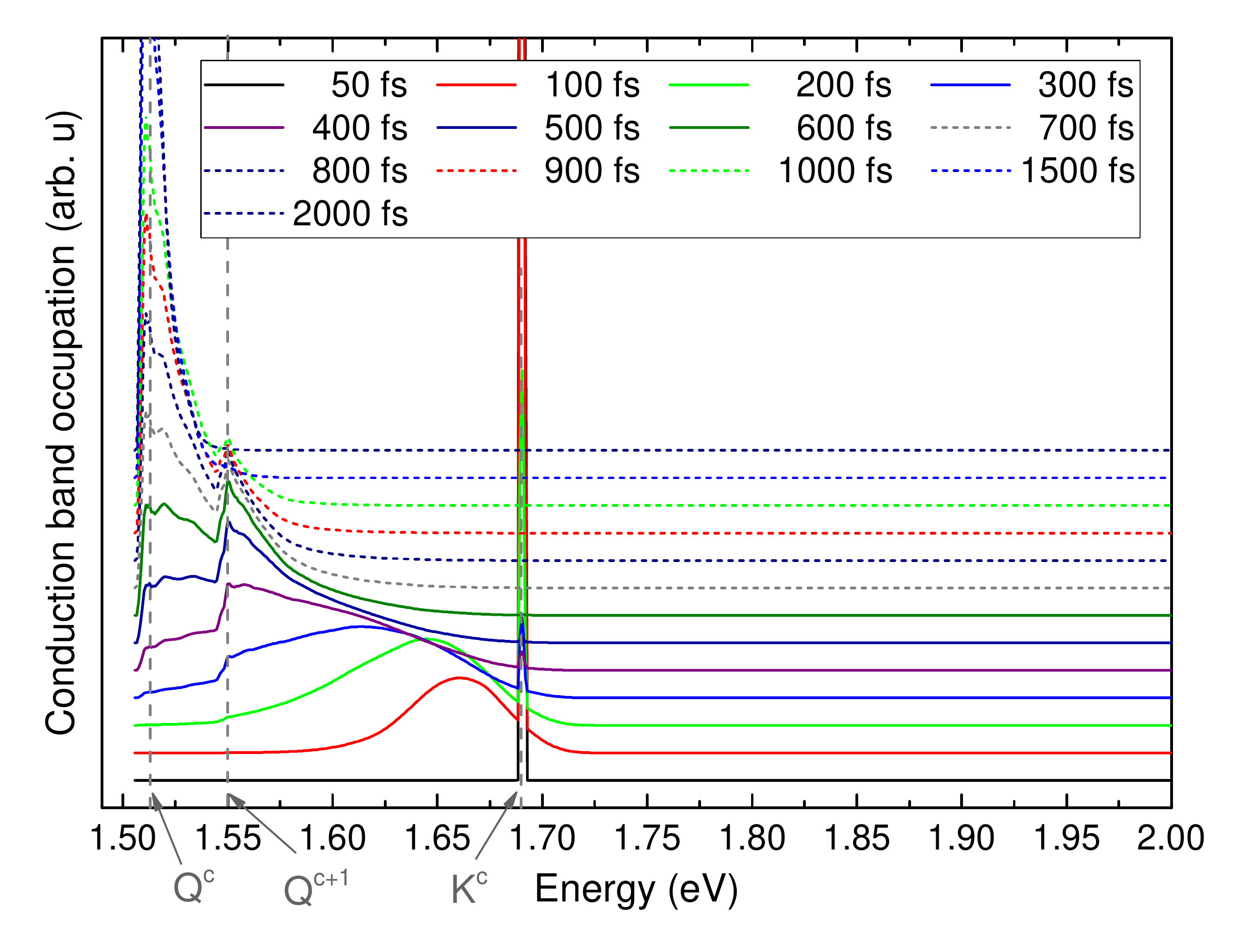}
\caption{\label{fig:MoSe2-WSe2_dynamics} (Color online) Time-evolution of the conduction band occupation of an AA'-stacked MoSe$_2$/WSe$_2$ heterostructure under the effect of electron-phonon scattering after an initial placement of 0.1 electrons in the MoSe$_2$-dominated conduction band minimum at each of the K and K' points ($K^{c}$) and a temperature of 10\,K. The diagram shows the energy-resolved evolution at selected points in time {\blau after the initial placement of the charge}. Contour plots of snapshots of the k-resolved sum of the band occupations over the Brillouin zone can be found in Section {\blau 5} of the supplemental material.}
\end{figure}
%
\begin{table*}
\caption{\label{tab:tab_gfactor} Calculated g-factors for selected momentum-direct and -indirect excitons. $g_{exc}$ and $g_{band}$ are the effective exciton g-factor as calculated from the excitonic wavefunctions and the g-factors in the two-bands approximation, respectively. Upper indices indicate the method used to calculate the interband transition energies entering Eq.~\ref{eq:morb}. All values are in units of $\mu_B$. }
\begin{ruledtabular}
\begin{tabular}{ c  c  c  c  c  c  c  c  c }
          &                    & $K^v\rightarrow K^c$ & $K^v\rightarrow K^{c+1}$ & $K^v\rightarrow K'^c$ & $K^v\rightarrow Q^c$ & $K^v\rightarrow Q^{c+1}$ & $\Gamma^v\rightarrow K^c$ & $\Gamma^v\rightarrow Q^c$ \\
\hline
AA        & $g_{band}^{DFT}$   & -5.95 ($\uparrow\uparrow$)      & -10.45 ($\uparrow\downarrow$) &  -16.88 ($\uparrow\downarrow$)   &  -8.41 ($\uparrow\uparrow$)  &  -12.16 ($\uparrow\downarrow$) &  7.14 ($-\uparrow$)                    &  4.68 ($-\uparrow$)      \\
          & $g_{exc}^{DFT}$    &  -5.89               &  -10.38                        &  -16.80           &                     &                           &                           &              \\
          & $g_{exc}^{GW}$     &  -5.00               &  -9.27                        &  -13.88           &                     &                           &                           &              \\
          & Exp                &   -8.5\footnotemark[1]\footnotetext{Ref.~{\blau[\onlinecite{Ciarrocchi-nature-MoSe2WSe2}]}}, -5.5$\pm$0.8\footnotemark[2]\footnotetext{Ref.~{\blau[\onlinecite{foerg2020moire}]}}            &                          &                   &                     &                           &              \\
\hline
AA'       & $g_{band}^{DFT}$   &  -16.48 ($\uparrow\downarrow$)      & -12.01 ($\uparrow\uparrow$)                   & -5.65 ($\uparrow\uparrow$)                  &  -9.15 ($\uparrow\uparrow$)             & -12.98 ($\uparrow\downarrow$)                  &  -6.94 ($-\uparrow$)                    &  0.4 ($-\uparrow$)           \\
          & $g_{exc}^{DFT}$    &  -16.47\footnotemark[3]\footnotetext{12x12x1 k-point grid}   & -12.01\footnotemark[3] & -5.65\footnotemark[3] & -9.14\footnotemark[3]                     & -12.82\footnotemark[3]\footnotemark[3] & -5.34\footnotemark[3]             &  -3.76\footnotemark[3]        \\
          &                    &  -16.34                    & -11.87                         &   5.55             &                    &                            &                           &  \\
          & $g_{exc}^{GW}$     &  -13.67\footnotemark[3]                          & -9.41\footnotemark[3] &  -4.83\footnotemark[3]              & -7.39\footnotemark[3]            & -11.00\footnotemark[3] & -4.35\footnotemark[3]                 &  -3.45\footnotemark[3]        \\
          &                    &  -13.58                 & -9.31                         &  4.75               &                    &                            &                           &              \\
          & Exp                & 15.2\footnotemark[4]\footnotetext{Ref.~{\blau[\onlinecite{Wang2020}]}, signs of g-factor not reported}, -16\footnotemark[5]\footnotetext{Ref.~{\blau[\onlinecite{Delhomme2020}]}}  & 10.7\footnotemark[4]                            &                   &                     &                           &              \\
\hline
AB        & $g_{band}^{DFT}$   &  -5.91 ($\uparrow\uparrow$)              & -10.43 ($\uparrow\downarrow$)                  & -16.88 ($\uparrow\downarrow$)            &  -8.11 ($\uparrow\uparrow$)           & -11.76 ($\uparrow\downarrow$)    &  6.10 ($-\uparrow$)                    &  3.90 ($-\uparrow$)      \\
          & $g_{exc}^{DFT}$    &  -5.81               &  -10.30                &  -16.74                 &                     &                           &              \\
          & $g_{exc}^{GW}$     &  -4.96               &  -9.23                &  -13.88                  &                     &                           &              \\
          & Exp                &  7.1\footnotemark[6]\footnotetext{Ref.~{\blau[\onlinecite{Ciarrocchi-nature-MoSe2WSe2}]}}                     &                          &                   &                     &                           &              \\
\end{tabular}
\end{ruledtabular}
\end{table*}

Clearly, an additional distinguishing factor between the 'direct' $K\rightarrow K$ and the 'indirect' $K\rightarrow Q$ and $\Gamma\rightarrow Q$ excitons should be their temperature dependence, {\blau as the momentum-indirect excitons require} the assistance of momentum sources for their formation and radiative recombination. In typical experiments, the excitation laser energy is too small for appreciable optical excitation at the $\Gamma$ and $Q$ points and instead optically excites electrons at the $K$ and $K'$ points. In order to obtain some insight into the scattering dynamics of optically excited electrons and the relevant timescales, we used the Boltzmann Transport Equation {\blau(BTE)} as implemented in the PERTURBO code~\cite{perturbo} to calculate the time evolution of the conduction band occupations of an AA' stacked MoSe$_2$/WSe$_2$ under the influence of electron-phonon coupling~\footnote{We used the Quantum ESPRESSO~\cite{qe} package to calculate the electronic structure and phonon spectra on a uniform grid of 9x9x1 k-points using the local density approximation and normconserving pseudopotentials from the PseudoDojo repository~\cite{pseudodojo}. The electronic wavefunctions were expanded in a planewave basis with a cutoff energy of 120\,Ry, allowing for well converged phonon frequencies. We included spin-orbit coupling in our calculations and corrected the electronic structure with the GW corrections to match the electronic bands shown in Figure~\ref{fig:geometry_bands}. Using the results of these results as input, we then used the PERTURBO code to calculate the temporal dynamics of excited electrons in the conduction bands. For this, we initially placed a charge of 0.1\,$e$ in the conduction band minimum at each of the K and K' points of the hexagonal Brillouin zone and propagated the band occupations through solution of the Boltzmann Transport Equation including electron-phonon coupling induced scattering for a simulation time of 2\,ps and {\blau temperatures of 10\,K and 300\,K}. To ensure an accurate representation of the scattering processes, a wannierization procedure was used to interpolate the electron-phonon matrix elements from a coarse 9x9x1 k-point grid to a dense 90x90x1 grid.}. Selected snapshots of the {\blau energy-resolved conduction band occupation for a temperature of T=10\,k} are shown in Figure~\ref{fig:MoSe2-WSe2_dynamics}. Our calculations suggest that, neglecting {\blau the Coulomb interaction} between electrons and holes, the electronic charge is efficiently scattered away from the $K$/$K'$ minima into the conduction band valley around the $Q$ point by phonons with momenta slight smaller than $\overline{KQ}$ within several 100\,fs. We find a similar thermalization timescale if the initial charge is placed into the higher energy conduction bands at the K/K' points that are contributed by the WSe$_2$ layer {[cf. Sec.~4 of the SI]}. Due to the similar exciton binding energies, it would be a reasonable conclusion from these results that there should be a significant population of the global conduction band minimum at the $Q$ point {\blau even at low temperatures}, and thus of the lowest-energy excitons over the fundamental bad gap, competing with radiative recombination of $\mathbf{Q}\approx$0 excitons at the $K$/$K'$ points of the heterostructure. 

On the other hand, both recent magneto-optical experiments~\cite{Ciarrocchi-nature-MoSe2WSe2,Wang2020,Delhomme2020,foerg2020moire} and theoretical studies~\cite{kunstmann-paper,foerg2020moire} suggest that the shift of interlayer exciton binding peak positions under external magnetic fields are a substantial factor for establishing the nature of the contributions to the interlayer excitonic peaks. It was found from magneto-luminescence measurements and similar techniques that the interlayer excitons of MoSe$_2$/WSe$_2$ heterostructures exhibit g-factors that (i) substantially differ from the g-factors of intralayer excitons in the monolayer TMDC materials (usually $\left|g\right|\approx 4$) and {\blau (ii)} show a strong dependence on the stacking order: While, for heterostructures in the R registry (e.g. {\blau the }AA, AB stacking orders in the convention used in this paper), g-factors between 4.2 and 8.5 were found~\cite{Ciarrocchi-nature-MoSe2WSe2,foerg2020moire}, the interlayer exciton g-factors of heterostructures in the H registry are substantially higher, 15-16 for the energetically lowest contribution~\cite{Wang2020,Delhomme2020}. Based on theoretical results from density functional theory calculations, the interlayer excitons have thus been interpreted to be zero-momentum singlet and triplet excitations at the $K$ and $K'$ points. 

The g-factor of a band transition between a valence band state $v$ with crystal momentum $\mathbf{k}$ and a conduction band state $c$ at $\mathbf{k}'$ can be calculated from the relation~\cite{deilmann-gfactor}
\begin{equation*}
g^{band}_{v\mathbf{k},c\mathbf{k}'} = m^z_{c\mathbf{k}'}-m^z_{v\mathbf{k}},
\end{equation*}
with the band magnetic moments (in units of $\mu_B$) of valence and conduction bands,
\begin{equation*}
m^z_{n\mathbf{k}}=m^{orb,z}_{n\mathbf{k}}+m^{spin,z}_{n\mathbf{k}}.
\end{equation*}
Here, $m^{spin,z}_{n\mathbf{k}}$ is the expectation value of spin momentum operator in out-of-plane direction
$m^{spin,z}_{n\mathbf{k}}=-\frac{eg_e}{2m_e\mu_B}\left\langle\Psi_{n\mathbf{k}}|\hat{S}_z|\Psi_{n\mathbf{k}}\right\rangle$,
with spinor wavefunctions $\Psi_{n\mathbf{k}}$. $m^{orb}_{n\mathbf{k}}$ is the corresponding orbital contribution to the magnetic moment and for a magnetic field in z-direction is given by the relation~\cite{roth-gfactor}
\begin{equation*}
\begin{split}
m^{orb,z}_{n\mathbf{k}} =-\frac{i}{m_e\mu_B}\sum_{j\neq n}\left(\frac{\left\langle\Psi_{n\mathbf{k}}|\hat{\pi}_x|\Psi_{j\mathbf{k}}\right\rangle\left\langle\Psi_{j\mathbf{k}}|\hat{\pi}_y|\Psi_{n\mathbf{k}}\right\rangle}{E_{j\mathbf{k}}-E_{n\mathbf{k}}}\right.\\
 \qquad-\left.\frac{\left\langle\Psi_{nk}|\hat{\pi}_y|\Psi_{j\mathbf{k}}\right\rangle\left\langle\Psi_{j\mathbf{k}}|\hat{\pi}_x|\Psi_{n\mathbf{k}}\right\rangle}{E_{j\mathbf{k}}-E_{n\mathbf{k}}}\right),\label{eq:morb}
\end{split}
\end{equation*}
where $\pi$ is given by the momentum operator plus possible contributions from spin-orbit coupling~\cite{kunstmann-paper}. Both $m^{orb}_{nk}$ and $m^{spin}_{nk}$ can be extracted in a {\blau straightforward} way from density functional theory calculations.
\begin{figure}
\centering
\begin{minipage}{0.49\columnwidth}
\includegraphics*[width=\columnwidth]{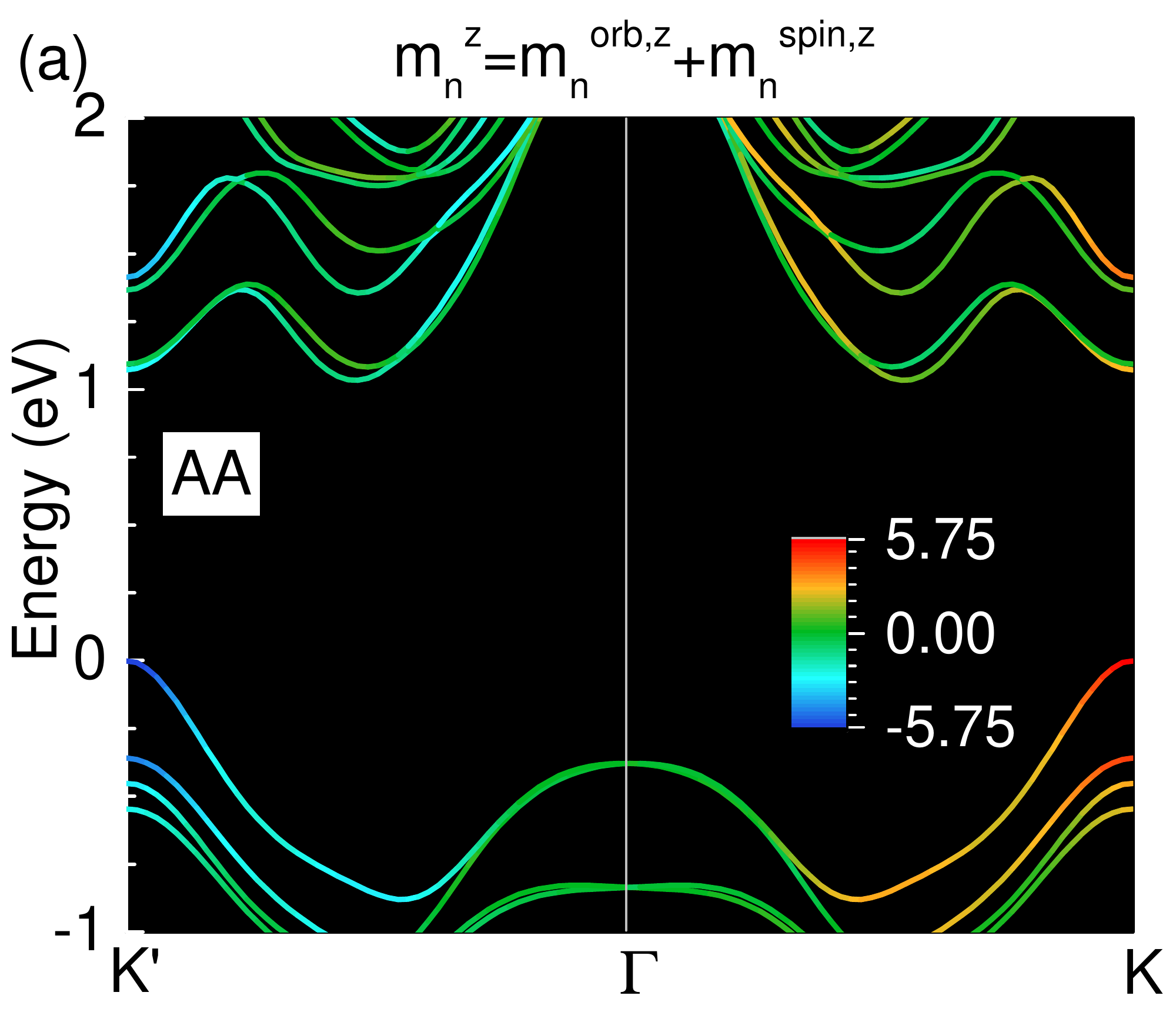}
\end{minipage}
\begin{minipage}{0.49\columnwidth}
\includegraphics*[width=\columnwidth]{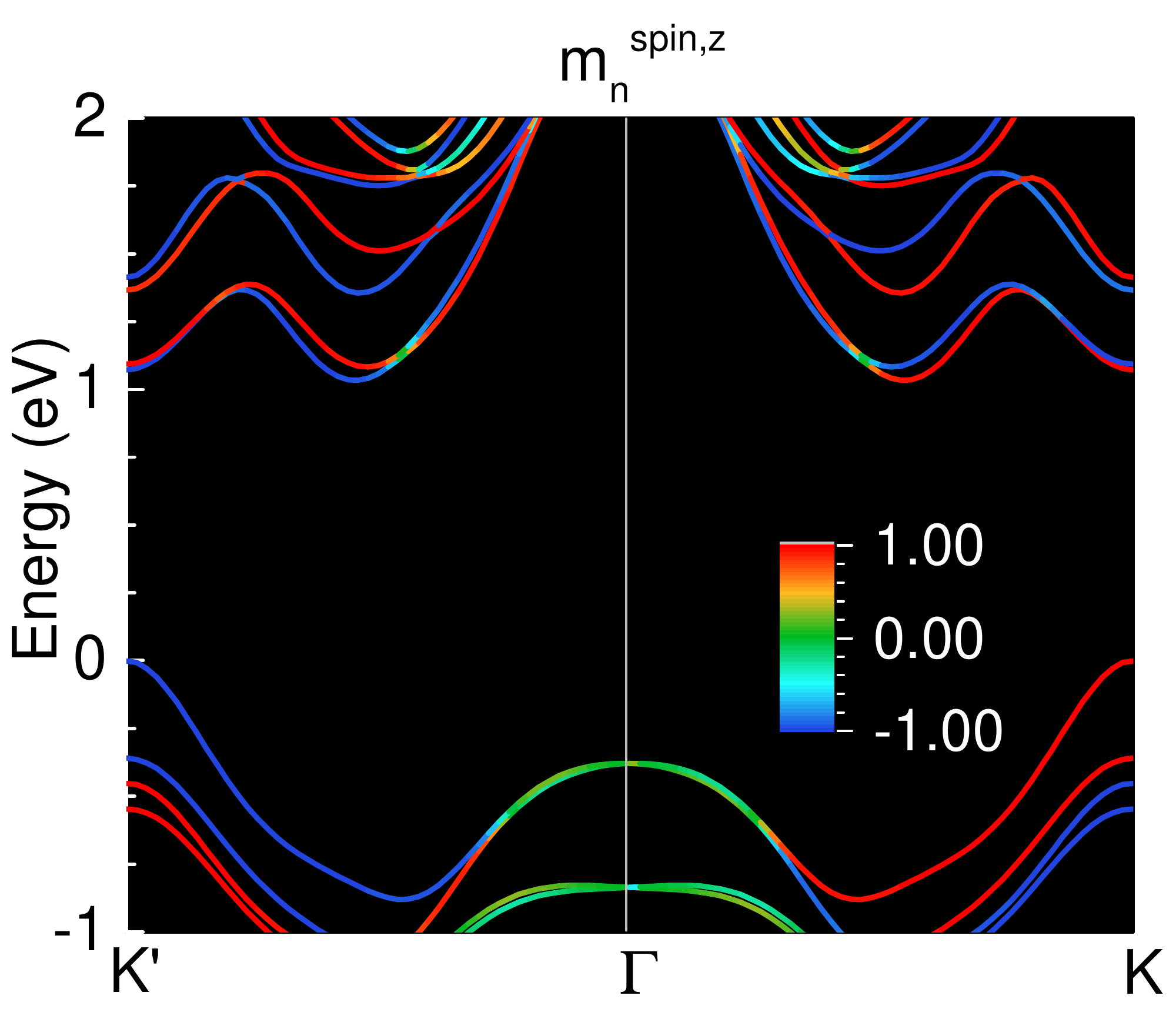}
\end{minipage}
\\
\begin{minipage}{0.49\columnwidth}
\includegraphics*[width=\columnwidth]{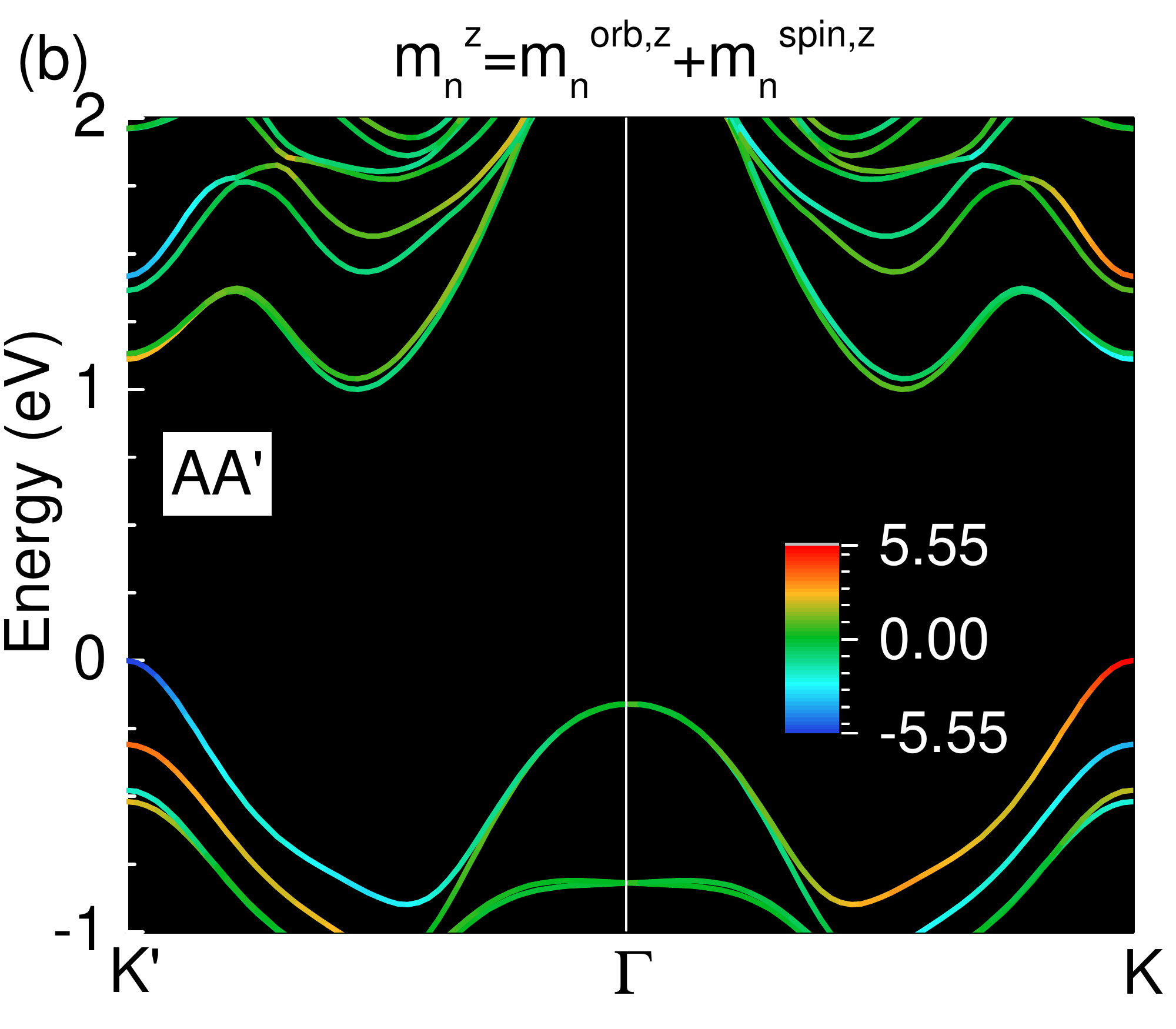}
\end{minipage}
\begin{minipage}{0.49\columnwidth}
\includegraphics*[width=\columnwidth]{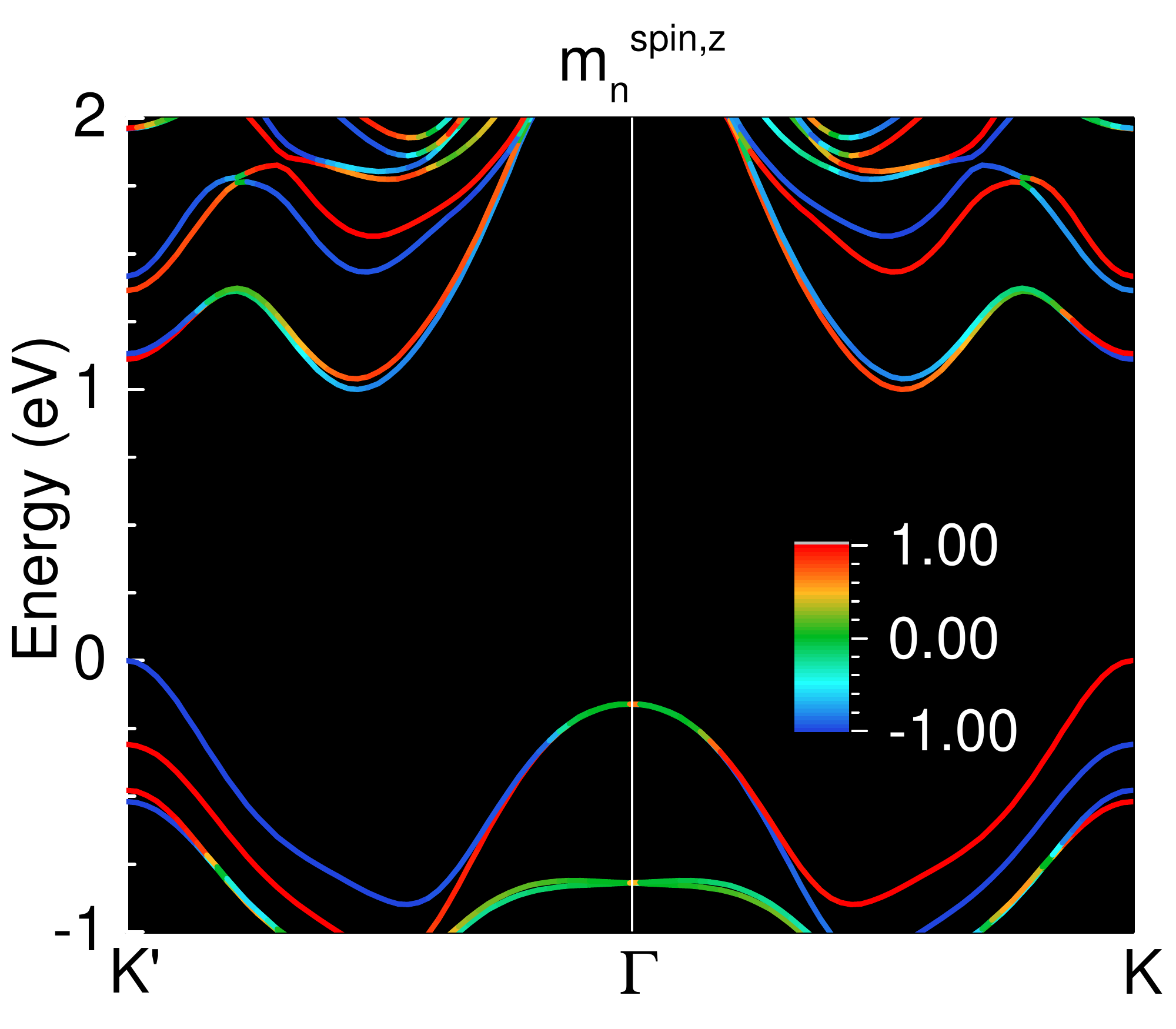}
\end{minipage}
\caption{\label{fig:mfactor} (Color online) Left side: Calculated band magnetic moments (in units of $\mu_B$) of valence and conduction bands of MoSe$_2$/WSe$_2$ with AA and AA' stacking order. Right side: Spin contributions to the band magnetic moments.}
\end{figure}

Within the typically implemented excitonic Bethe-Salpeter Equation formalism, the exciton is described by a weighted superposition of different interband transitions. With this, one can define an effective g-factor for an excitonic state $S$ with center-of-mass momentum $\mathbf{Q}$ by
\begin{equation*}
g^{exc}_{S,\mathbf{Q}}=2\sum_{v,c,\mathbf{k}}A_{vck}^{S,\mathbf{Q}}g^{bands}_{v\mathbf{k},c\mathbf{k}+\mathbf{Q}},
\end{equation*}
where the excitonic wavefunction $A_{vk,ck'}$ describes the weight of each transition contributing to the exciton. Depending on the exciton in question, the mixing of different transitions in the BSE approach might thus lead to somewhat different $g^{exc}$ compared to $g^{band}$ for a given excitonic transition. For instance, Deilmann~\emph{et al.} reported recently{\blau~\cite{deilmann-gfactor}} that even for the $A$ excitons of monolayer TMDCs, which are very localized in reciprocal space and consist to more than 95\,\%  of a $v\rightarrow c$ transition at the K/K' points, the additional small contributions can lower $g^{exc}_A$ compared to $g^{band}_{cvK}$ by about 30\,\%.
\begin{figure*}
\centering
\includegraphics*[width=0.9\textwidth]{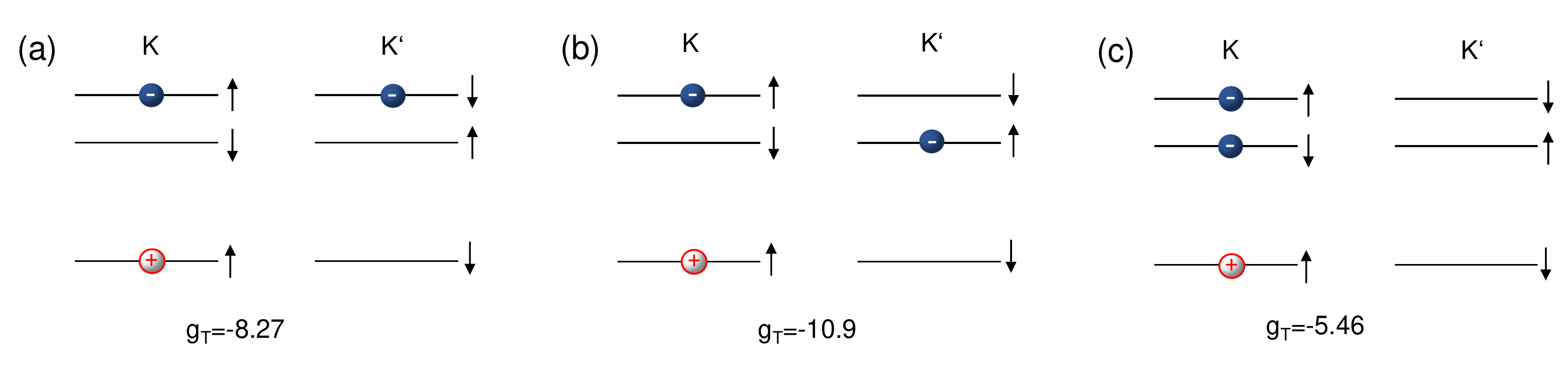}
\caption{\label{fig:trion-gfactor} \blau (Color online) Electron and hole configurations of three selected negatively-charged trions associated with the 'bright' interlayer exciton $Y_0$ in a AA'-stacked MoSe$_2$/WSe$_2$ heterostructure. In all cases, a radiatively combining electron-hole pair is located at the K point of the hexagonal Brillouin zone with the excess charge located either in the K or K' valleys. The calculated effective g-factor $g_T$ is given for each configuration.}
\end{figure*}

Table~\ref{tab:tab_gfactor} shows the calculated effective exciton g-factors $g^{exc}$ and band approximations $g^{band}$ for selected momentum-direct and -indirect excitons of an AA' stacked MoSe$_2$/WSe$_2$ heterostructure\footnote{The values for the indirect excitons were obtained from calculations with a rather coarse 12x12x1 k-point grid, but explicit inclusion of spin-orbit coupling in the BSE calculations.}. Plots of the underlying magnetic moments of valence and conduction bands and of the spin contribution for AA- and AA'-stackings are depicted in Figure~\ref{fig:mfactor}. Our calculated values for $g^{band}$ are in reasonable agreement with previous theoretical studies and the experimental data and support the assignment of the interlayer excitons to a pair of momentum-direct singlet and triplet excitations at the $K$ and $K'$ points. While the g-factors of the energetically lowest $K\rightarrow Q$ transitions show a certain dependence on the stacking order, this dependence appears to originate in changes in interband hybridization at the $Q$ point. The induced variation of the g-factors and energetic order of the exciton with the higher and the exciton with the lower g-factor are not consistent with the available experimental data. A similar conclusion can be reached for the other momentum indirect excitons. 

As expected from the previous reported results for the excitonic g-factors of the monolayer materials~\cite{deilmann-gfactor}, the inclusion of excitonic effects modifies the calculated effective g-factors {\blau of the MoSe$_2$/WSe$_2$ heterostructure}. For the $K\rightarrow K$ transitions and using the DFT band energies for the calculation of $m^{orb}$, the calculated $g^{exc}$ are about 1\,\% smaller than the corresponding $g^{band}$ values, which validates the band approximation for description of the momentum-direct $K$ interlayer excitons. We note that when using a coarser 12x12 k-point grid for the solution of the BSE, the difference to $g^{band}$ is even smaller, due to the larger weight put on contributions at the $K$ and $K'$ points in this case. The contributions to the indirect excitons are less localized in reciprocal space compared to the momentum-direct interlayer excitons and the differences between $g^{exc}$ and $g^{band}$ are larger in this case. A further change appears if we use GW band energies for the calculation of $m^{orb}$: due to the larger energy differences, the calculated g-factors are substantially smaller than those calculated from DFT-level band energies. This somewhat decreases the agreement between the experimental data and the predicted g-factors for the $K\rightarrow K$ excitons. The discrepancy might arise from the neglect of substrate and environmental effects in the simulations, affecting both the electronic band energies in the material and somewhat delocalizing the excitonic wavefunctions due to screening effects compared to the 'free-standing' material in vacuum. Another possible source of discrepancy is the use of DFT spinors for the calculation of the orbital magnetic moments, which is not fully consistent with the use of G$_0$W$_0$ transition energies.

{\blau As an alternative to the interpretation of the doublet structure of the interlayer PL peak
as a pair of charge-neutral excitons, Calman \emph{et al.}~\cite{geim-MoSe2WSe2} have recently suggested a pair of spatially indirect neutral exciton and spatially indirect negatively charged trion as the origin. A formalism for the estimation of the effective g-factor of trions was proposed and used by Lyons \emph{et al.} for the interpretation of the trion valley Zeeman splitting in monolayer WSe$_2$~\cite{lyons-trions}. For a specific negatively charged trion configuration defined by the valley indices $\tau$ (+1 for the K point valleys, -1 for the K' point valleys), the Zeeman splitting can be expressed through the relation
\begin{equation}
\Delta E=g_T\mu_BB_z = \frac{1}{2}\left(\tau_{exc}g_{exc}-\tau_cg_e-2g_l\right)\mu_BB_z\label{eq:gt},
\end{equation}
where $g_T$ is the trion effective g-factor due to valley- and spin-induced magnetic momenta, $g_{exc}$ and $g_e=2m_c^z$ are the exciton effective g-factor of the recombining electron-hole pair and the g-factor of an excess electron in a specific MoSe$_2$-dominated conduction band valley, respectively. The effect of the recoil of the excess electron after recombination of the electron-hole pair on the Zeeman splitting is included through a Landau level associated g-factor $g_l=2m^*_X/\left(m^*_cm^*_T\right)$.
%
Here, $m^*_e$ is the effective mass of the excess electron, $m^*_X=m^*_{v}+m^*_{c}$ is the exciton effective mass, and $m^*_T=m^*_X+m_e$ is the trion effective mass. We derived the effective masses of the valence band maximum and the spin-orbit split conduction band valley from our calculated DFT bandstructures [cf. Section 2 of the SI]. This results in $g_l$ in the range of 1.8-2.2 for all considered trion configurations, very similar to the experimentally deducted value in monolayer WSe$_2$~\cite{lyons-trions}. 

Based on Eq.~\ref{eq:gt}, we calculated the trion effective g-factors $g_T$ for the three trion configurations shown in Figure~\ref{fig:trion-gfactor} for the case of a AA'-stacked heterostructure. All three considered interlayer trion configurations consist of the bright neutral interlayer exciton $Y_0$ and an excess electron in different conduction band valleys and have been predicted to occur in the optical absorption spectra of vertically stacked MoS$_2$/WS$_2$~\cite{deilmann-MoS2WS2}. The trion binding energies of 25-28\,meV relative to the neutral interlayer exciton obtained in Ref.~{\blau[\onlinecite{deilmann-MoS2WS2}]} for the MoS$_2$/WS$_2$ heterostructure are similar to the trion binding energies predicted for TMDC monolayers~\cite{Berkelbach-trions}. Under the assumption of a similar binding energy in MoSe$_2$/WSe$_2$ heterostructures, all three trion configurations would give rise to absorption/emission peaks close to the 'darkish' $X_0$ interlayer exciton. Our results suggest that the three configurations should exhibit characteristic g-factors in the range of 5-11, distinctively different from the g-factor of 16.5 predicted for the $X_0$ interlayer exciton. 

Finally, we note that in principle, the exchange interaction between the three bound particles in the spin-triplet trion configuration shown in {\blau Figure}~\ref{fig:trion-gfactor}~(b) should give rise to an exchange-induced Berry curvature $\Omega_{ex}$~\cite{lyons-trions,Yu2014}, inducing an additional magnetic momentum and a g-factor contribution $g_{BP}=\frac{m_0}{2\hbar^2}\delta_{exch}\Omega_{ex}\left(\mathbf{k}=\mathbf{K}\right)$~\cite{lyons-trions}  ($\delta_{ex}$ is the exchange splitting, $m_0$ is the free electron rest mass). For monolayer WSe$_2$, this g-factor contribution has been estimated to be about $g_{BP}=$4, i.e. rather sizeable~\cite{lyons-trions,Yu2014}. However, due to the significant charge separation of electrons and holes in the case of the $K\rightarrow K$ interlayer excitons and the associated trions, and thus the marginal spatial overlap between the electron and hole wavefunctions (about 1/1000th of the spatial overlap of the electron and hole wavefunctions of the $Mo_A$ and $W_A$ intralayer excitons), we estimate the g-factor contribution induced by electron-hole exchange to be negligibly small compared to the situation in the monolayer materials.
}

\subsection{C-type excitons in AA'-stacked MoSe$_2$/WSe$_2$ heterostructures}
\begin{figure}
\centering
\includegraphics*[width=0.9\columnwidth]{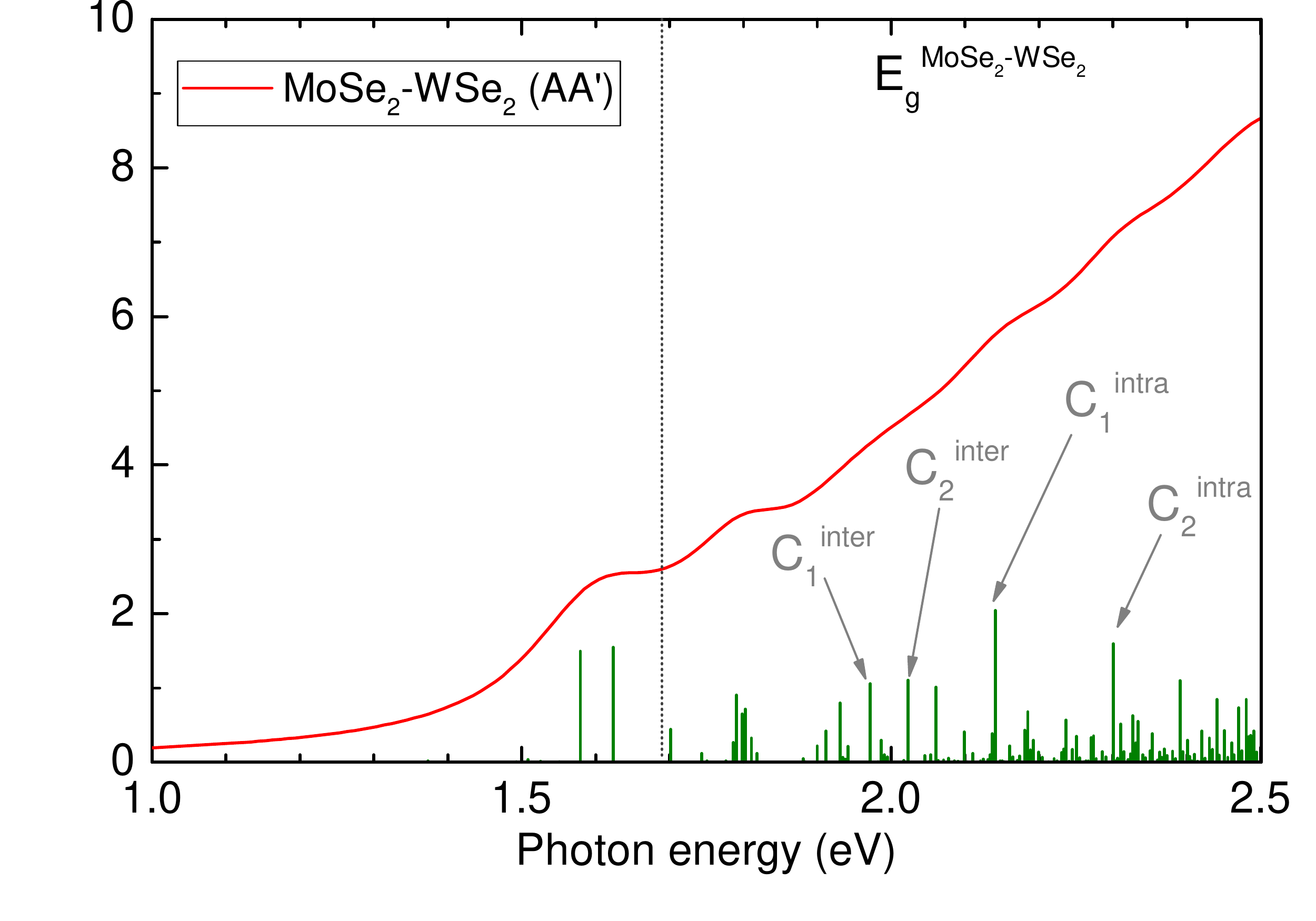}
\caption{\label{fig:MoSe2-WSe2_abs_C} (Color online) Calculated absorption spectrum of an AA' stacked MoSe$_2$/WSe$_2$ hetero\-structure. The labels indicate the positions of the four selected C-type excitons discussed in the main text.}
\end{figure}

In this section, we will go beyond the excitons around the absorption onset and consider the composition of the optical absorption at slightly larger energies. As we showed in section~\ref{sec:sec2}, the absorption spectra of monolayer molybdenum and tungsten based TMDCs feature strongly bound {\blau 'C'} excitons, which arise from a high joint-density-of-states between valence and conduction bands. The fulfillment of these band nesting conditions depend on details in the electronic structure. At the same time, the electronic contribution to the wavefunction of the monolayer C excitons feature a noticeable component that points in out-of-plane direction, which we found gave rise to a significant spatial interlayer delocalization of the excitonic wavefunctions for TMDC homobi- and trilayers. An interesting question is then how the tightly bound C excitons are affected by formation of a heterostructure. 

\begin{figure}
\centering
\includegraphics*[width=0.44\columnwidth]{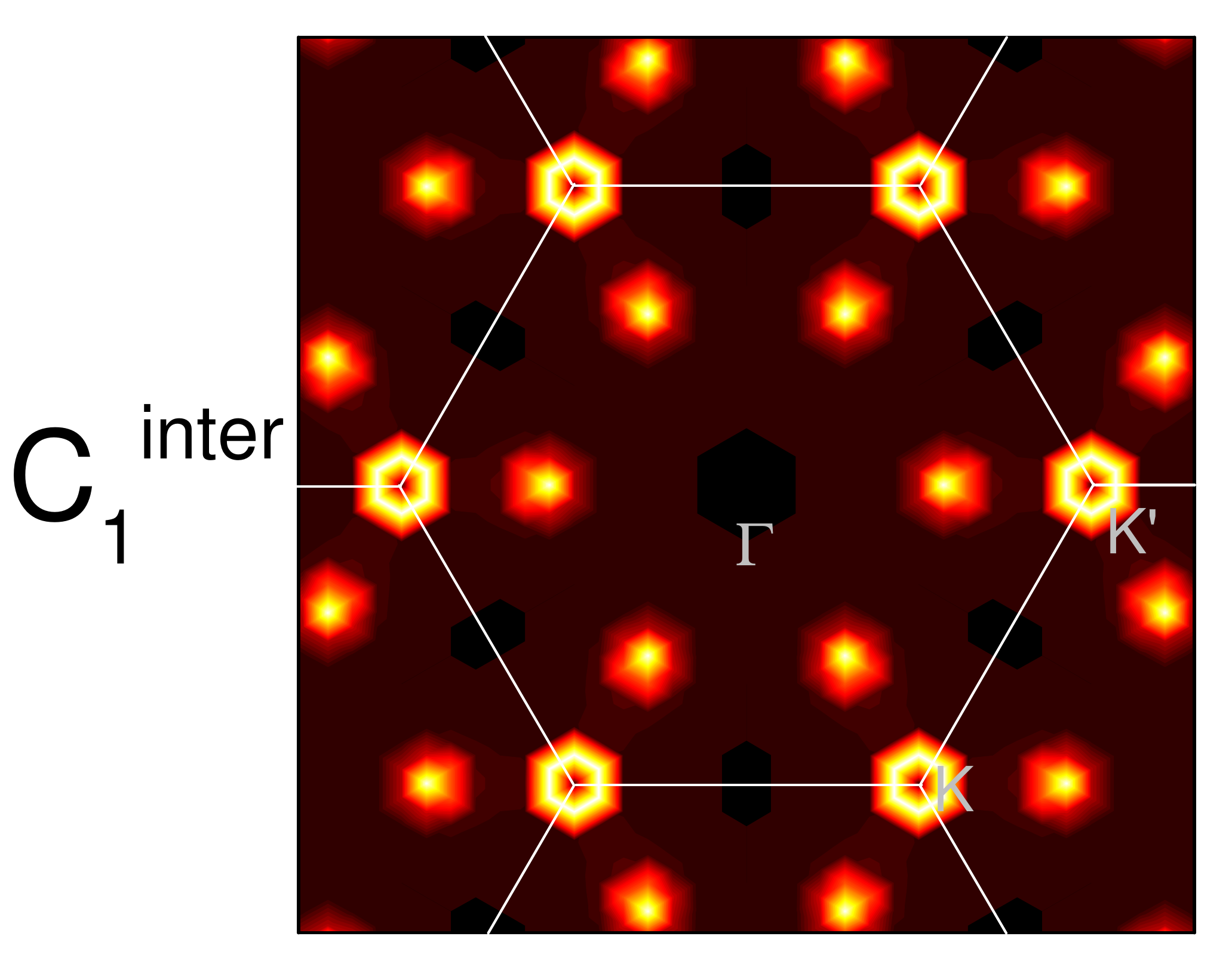}
\includegraphics*[width=0.44\columnwidth]{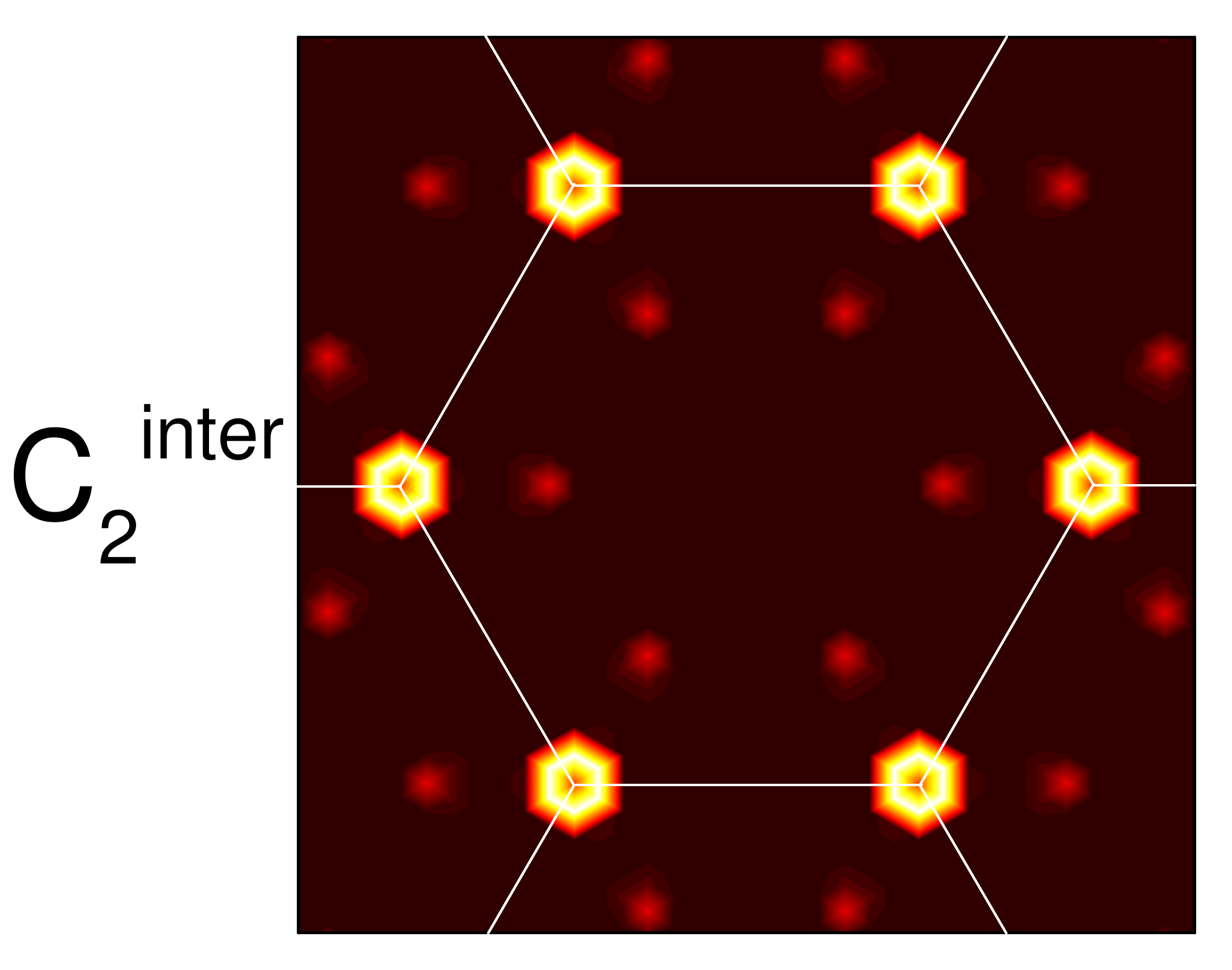}
\includegraphics*[width=0.44\columnwidth]{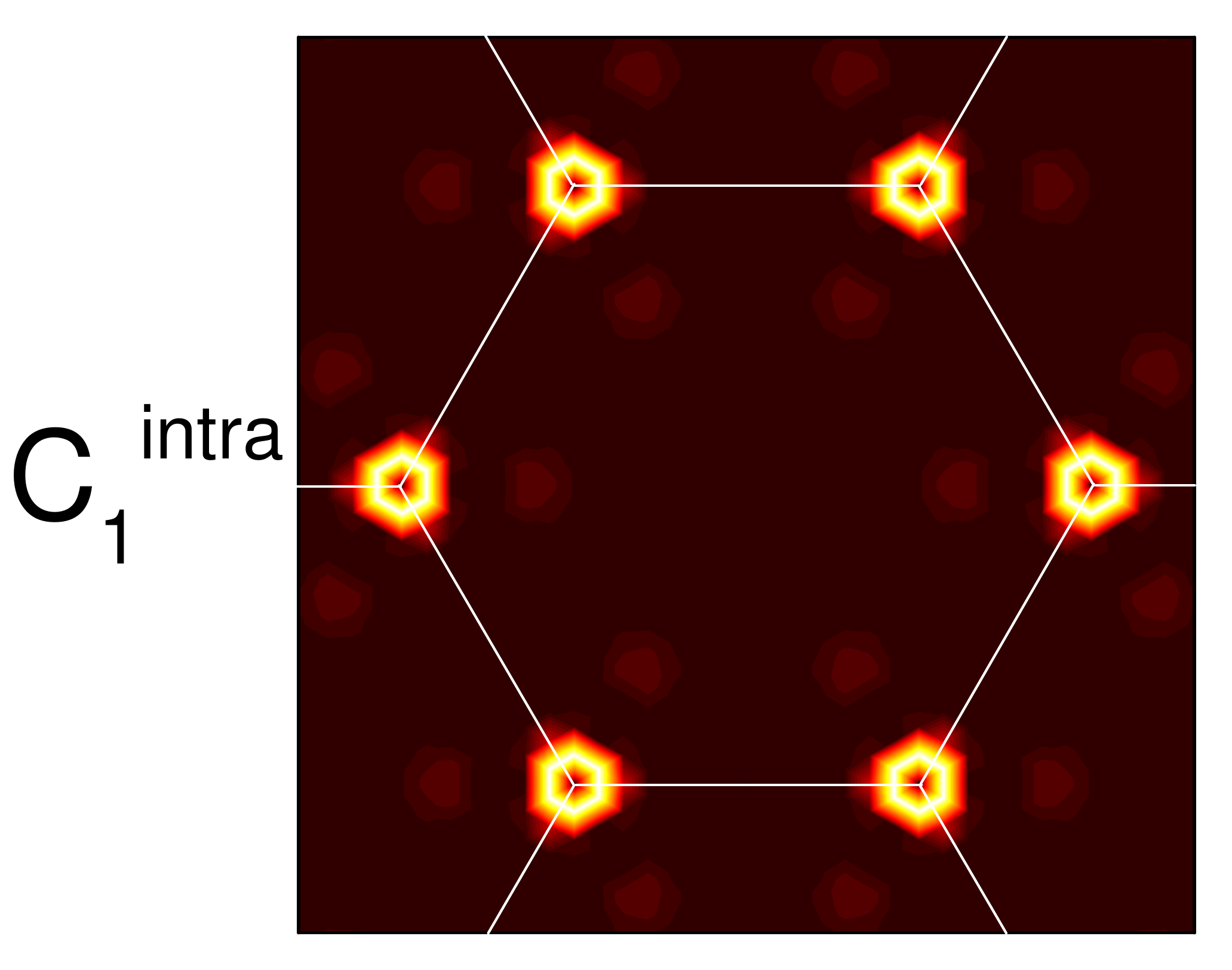}
\includegraphics*[width=0.44\columnwidth]{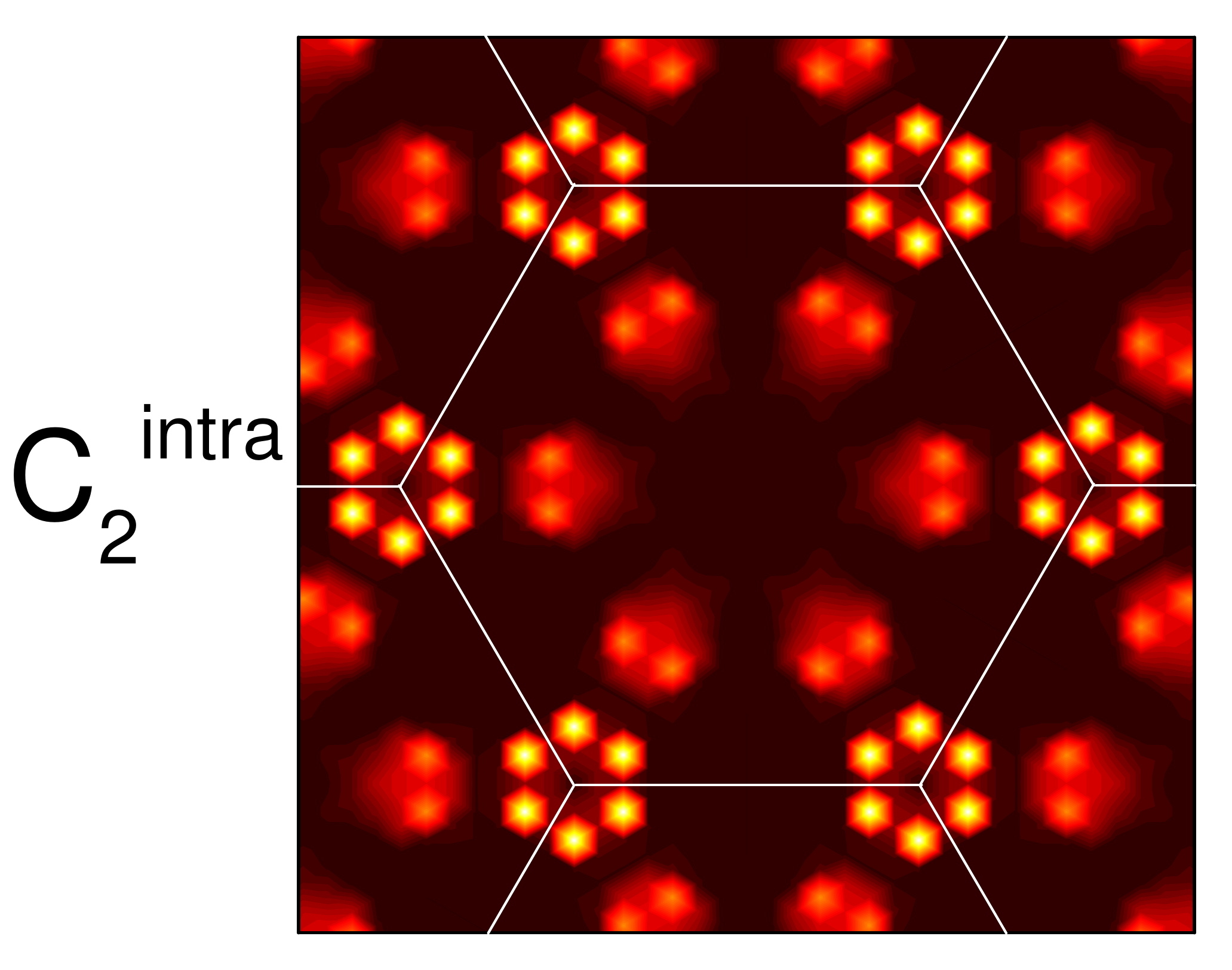}
\caption{\label{fig:MoSe2-WSe2_C_Ak} (Color online) k-resolved representation of the excitonic wavefunctions of four selected C-type excitons in AA stacked MoSe$_2$/WSe$_2$ heterostructures. Brighter colours indicate a larger contribution of direct band transitions at a given point in the Brillouin zone to the excitonic state.}
\end{figure}

The calculated optical spectrum of a AA' stacked MoSe$_2$/WSe$_2$ heterostructure shows a number of peaks with higher oscillator strength in the energy range of 2.0-2.5\,eV, where the C excitons of monolayer MoSe$_2$ and WSe$_2$ are located. In the following, we will focus on four of these bright transitions, as labeled in Figure~\ref{fig:MoSe2-WSe2_abs_C}. These states have in common that their k-resolved excitonic wavefunctions, i.e. the location of the band transitions contributing to the excitonic state in the Brillouin zone of the material, are very similar to those of the C excitons in monolayer MoSe$_2$ and WSe$_2$. Figure~\ref{fig:MoSe2-WSe2_C_Ak} shows the calculated k-resolved exciton wavefunctions. In all four cases, the strongest contributions come from the direct vicinities of the K and K' points, with additional contributions from a 'special' point on the $Q-K/K'$ lines. As for the monolayer materials, the location of this special point is given by a band nesting between valence and conduction bands, which appears to be present in the slightly modified electrionic bandstructure of the heterostructure as well. This suggests that all four selected absorption peaks are 'C-like'. 

\begin{figure}
\centering
\begin{minipage}{0.49\columnwidth}
\includegraphics*[width=\columnwidth]{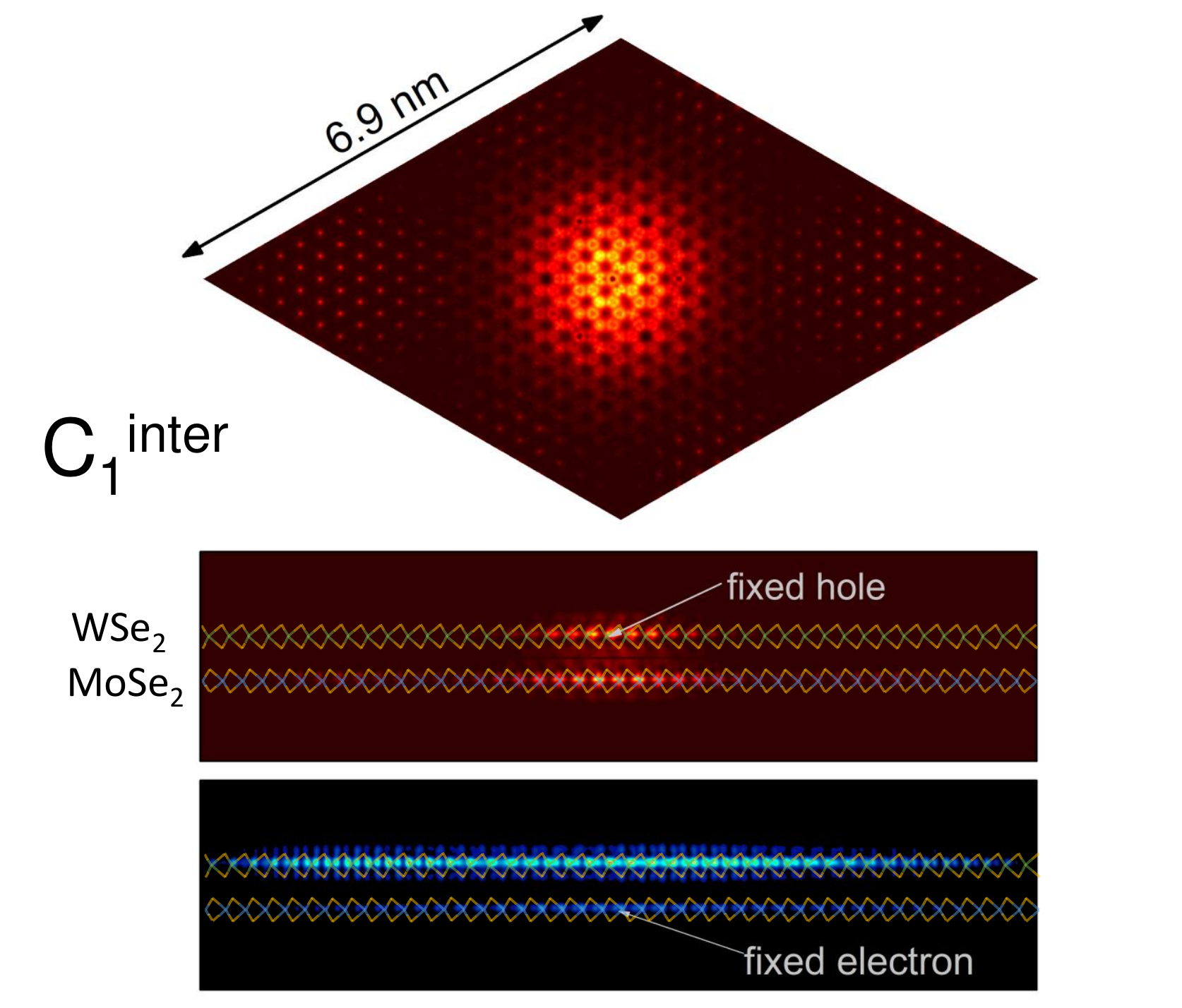}
\end{minipage}
\begin{minipage}{0.49\columnwidth}
\includegraphics*[width=\columnwidth]{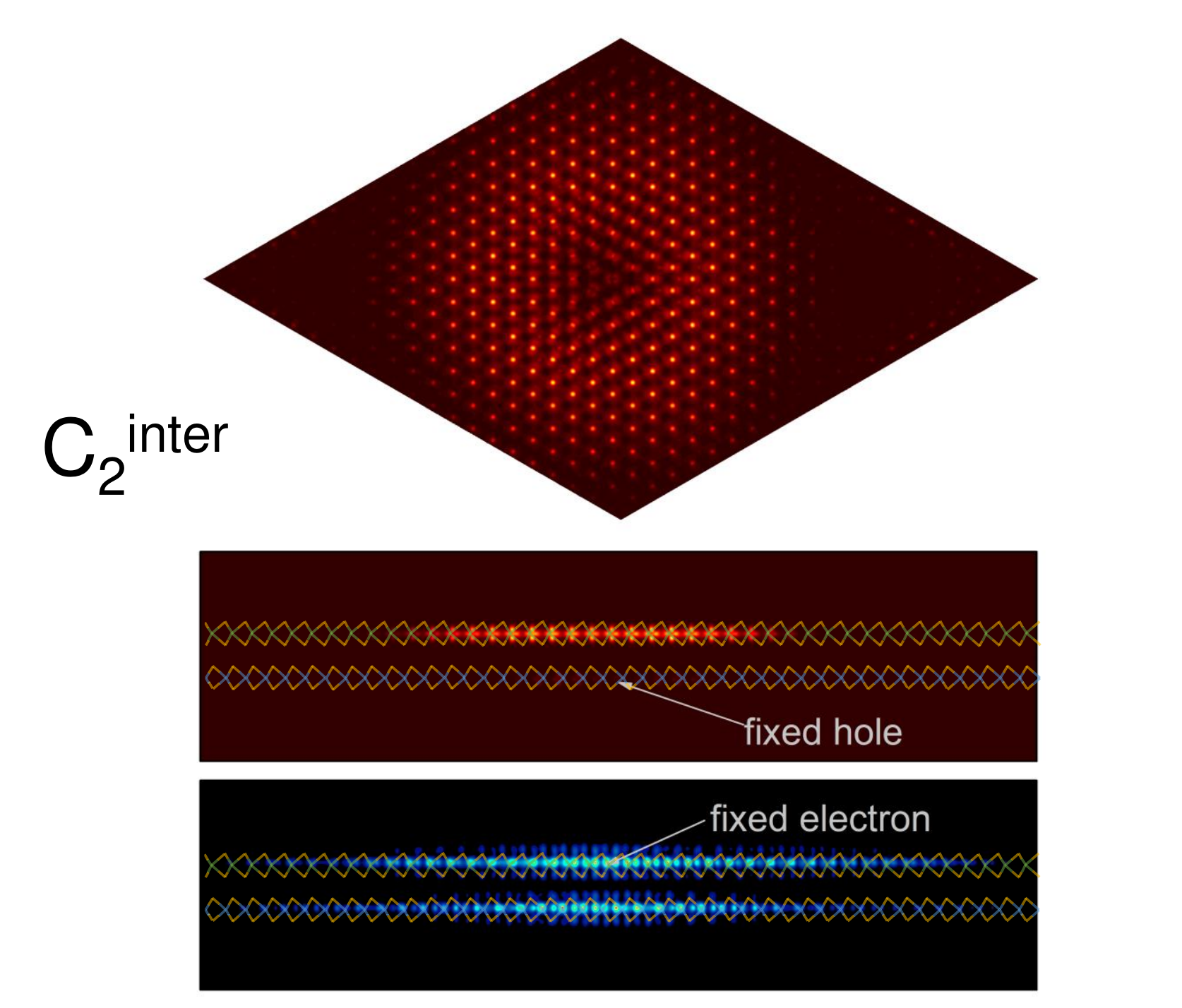}
\end{minipage}
\\
\begin{minipage}{0.49\columnwidth}
\includegraphics*[width=\columnwidth]{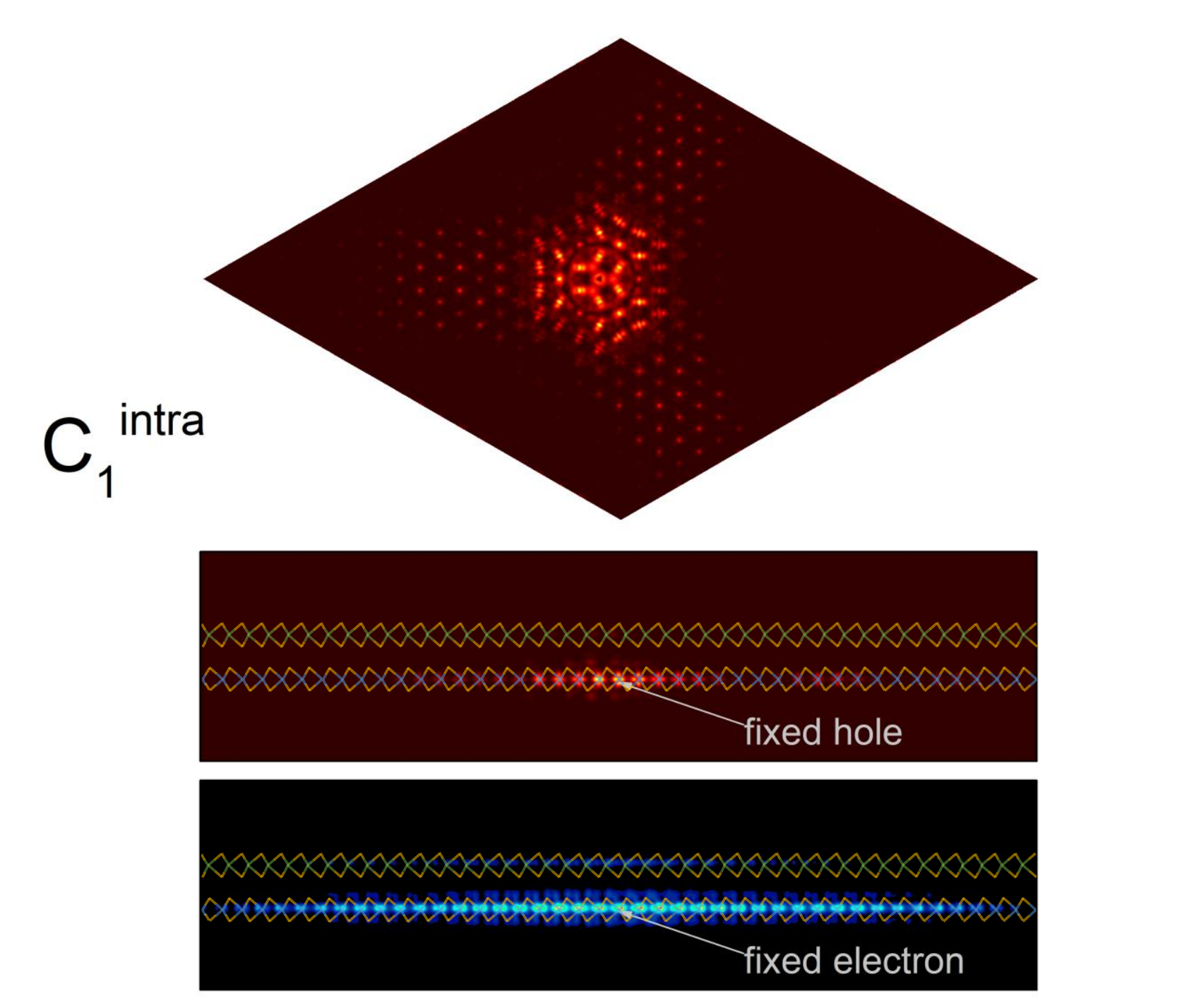}
\end{minipage}
\begin{minipage}{0.49\columnwidth}
\includegraphics*[width=\columnwidth]{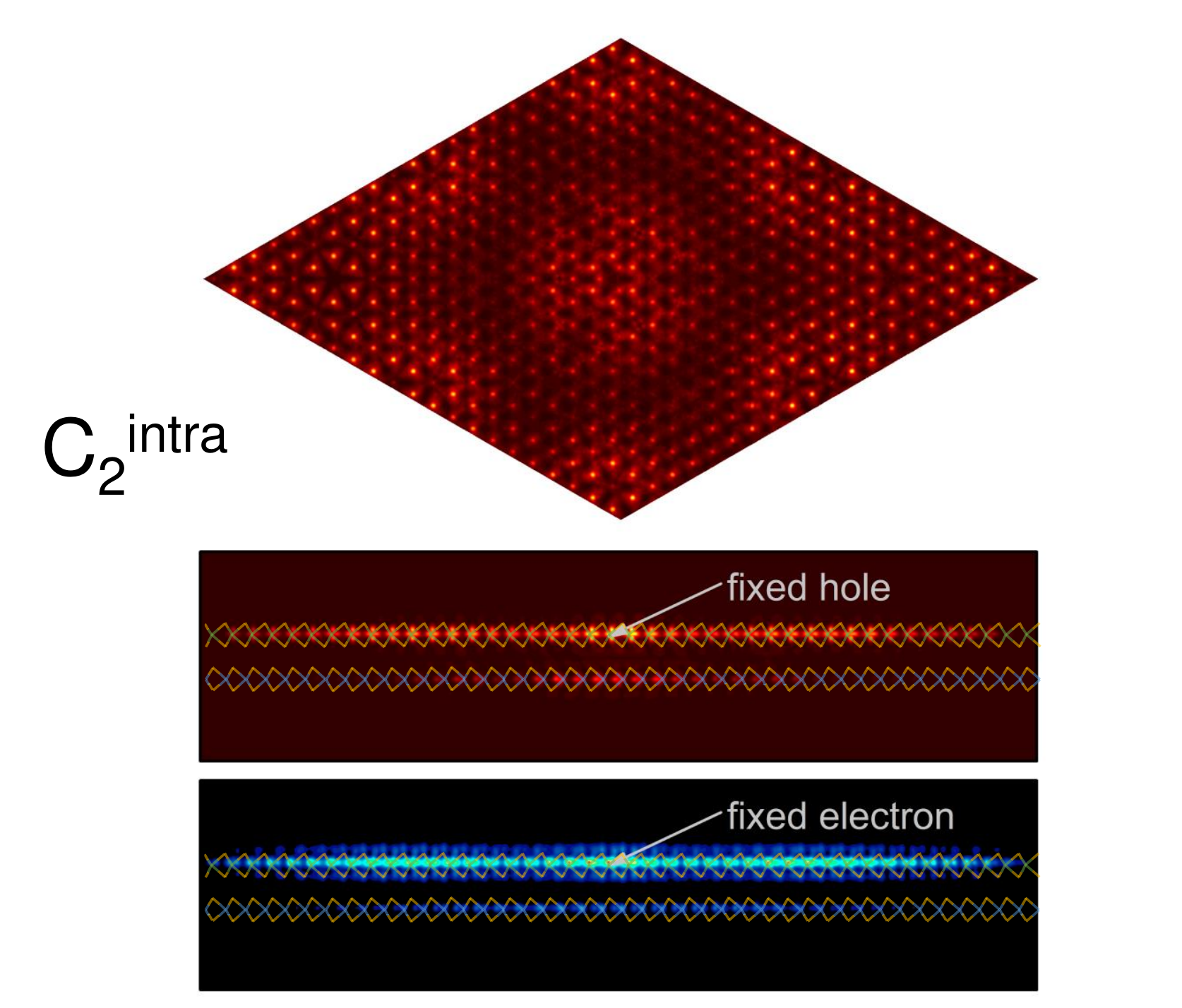}
\end{minipage}
\caption{\label{fig:MoSe2-WSe2_C_wfc} (Color online) Plots of the electron (red) and hole (blue) parts of the excitonic wavefunctions of four C-like excitons in a AA'-stacked MoSe$_2$/WSe$_2$ heterostructure. Top parts of the subfigures show projections of the wavefunction onto the x-y plane, bottom parts show projections onto the x-z plane.}
\end{figure}

As the plots of the real-space excitonic wavefunctions in Figure~\ref{fig:MoSe2-WSe2_C_wfc} show, we can divide the four selected transitions into two groups: Two transitions at lower energies, $C_1^{inter}$ and $C_2^{inter}$, show a distinct interlayer nature. $C_1^{inter}$ arises from transitions from the WSe$_2$-dominated valence band into the MoSe$_2$-dominated lowest conduction band, which, interestingly, suggests that the slope of these two bands is sufficiently similar to establish an effective band nesting condition even between bands contributed by different materials. The corresponding excitonic wavefunction shows a clear confinement of the hole part to the WSe$_2$ layer, while the electron part is mostly located in the MoSe$_2$ layer, but spills over into neighboring WSe$_2$ layer, similar to the exciton wavefunction in MoSe$_2$ homobilayers. The electronic part of the exciton wavefunction shows a bright and localized 'core' with some additional density tail from band-like states of similar energy that are mixed into the exciton wavefunction. This localization is also mirrored in the exciton binding energy: based on the energy difference between the highest valence and the lowest conduction band at the special point, which our calculations indicate to have the largest contribution to the exciton, we estimate the exciton binding energy to be about 453\,meV. 

The situation is somewhat reversed for $C_2^{inter}$: here, the largest contributions come from transitions at the K and K' points and their immediate vicinity, between the second and third highest valence bands (of purely MoSe$_2$ or WSe$_2$ nature) into the third lowest conduction band, which is contributed by the WSe$_2$ layer, see the electronic bandstructure shown in Figure~\ref{fig:geometry_bands}. Smaller contributions come from transitions at the band nesting point, between the valence band top and the hybridized second lowest conduction band. Consequently, it is the hole part of the excitonic wavefunction that shows a distinct interlayer delocalization in this case, with the larger share of the hole being located in the MoSe$_2$ layer, while the electronic part is well{\blau-}localized in the WSe$_2$ and only shows a minor spilling into the neighboring material. Within the layer, the electron appears to be well localized, with an exciton radius of about 35\,\AA. Based on the energy difference between the initial and final states at the $K$ point and the exciton peak position, we estimate the exciton binding energy to be 448\,meV. We can thus understand $C_1^{inter}$ and $C_2^{inter}$ as 'interlayer C excitons', with a noticeable interlayer separation of bound electrons and holes.

In contrast to the previous two states, the second group of transitions can be more or less clearly be attributed to arise from individual layer and thus have a noticeable intralayer nature. $C_1^{intra}$ consists of a mix of almost degenerate MoSe$_2\rightarrow$WSe$_2$ transitions close to the $K$ and $K'$ points and additional C-like transitions away from $K$. Despite this, we find that the exciton wavefunction has a significant intralayer nature centered on the MoSe$_2$ layer and the electron part is quite localized. In terms of both spatial distribution {\blau and contributions} in reciprocal space, $C_1^{intra}$ appears to be analogous to the C excitons in mono- and bilayer MoSe$_2$. From a weighted average of the energies of the transitions contributing to the exciton wavefunctions, the binding energy is calculated to be 400\,meV. The $C_2^{intra}$ transition, on the other hand, mainly consists of intralayer transitions within the WSe$_2$ layer and thus has some resemblance to the C excitons of WSe$_2$. Quite surprisingly, while the electron and hole parts of the exciton wavefunction appear to be quite delocalized, we calculate a rather high exciton binding energy of 414\,meV from the difference of the peak energy to a weighted average of the energies of the constituing band transitions. It is thus possible that the calculated $C_2^{intra}$ peak in fact consists of an exciton mixed with band transitions of similar energy, which gives rise to an extended wavefunction. A more detailed decomposition would be necessary to draw further conclusions on the nature of the $C_2^{intra}$ peak.

\section{Summary}
We have {\blau expanded} on our previous study on the excitonic spectra of vertically stacked MoSe$_2$/WSe$_2$ heterostructures. The previously reported significantly stacking-dependent optical oscillator strengths of the lowest energy interlayer excitons lead to a wide range of radiative lifetimes. In particular, deriving an equation for excitons emitting out-of-plane polarized light, we were able to quantify the radiative lifetime of the {\blau'}brightish' interlayer excitons in AB-stacked heterostructures to 200\,ns (at T=4\,K), significantly larger than the lifetimes obtained for the investigated AA and AA' stacking orders. We expect these results to be transferable to other stacking orders as well.

By virtue of implementation and solution of the finite-momentum Bethe-Salpeter Equation, we further studied the binding energies and wavefunctions of relevant momentum indirect excitons between the valence and conduction band valleys. Here, we showed from ab initio that the momentum-indirect excitons over the fundamental band gap ($K\rightarrow {\blau Q}$) {\blau should be} the lowest-energy excitons {\blau for ideally stacked MoSe$_2$/WSe$_2$ heterostructures, about 0.15\,eV below the momentum-direct $K\rightarrow K$ excitations with a significant decrease of spatial charge separation due to interlayer hybridization effects}. {\blau Experimental observation of such momentum-indirect interlayer excitons have been reported recently by Hanbicki~\emph{et al}~\cite{hanbicki-MoSe2WSe2}.} Simulations of the carrier dynamics of free electrons suggest that the global conduction band minimum should be quickly populated after optical excitation in resonance with the intralayer 'A' excitons. Despite these results, our calculations of the effective exciton Land\'e g-factors confirm previous reports that the {\blau experimentally observed interlayer exciton magnetoluminescence in bilayer MoSe$_2$/WSe$_2$ heterostructures} probably stems from {\blau (quasi-)momentum direct} $K\rightarrow K$ {\blau interlayer} excitons. 
Particularly compared to the TMDC homobilayers and the related MoS$_2$/WS$_2$ heterostructures this raises the question why the $K\rightarrow Q$ excitons {\blau do not appear to be commonly observed in photoluminescence measurements of MoSe$_2$/WSe$_2$ bilayer heterostructures. Future theoretical and experimental work might shed further light on, e.g. the influence of the moir\'e potential on the interlayer excitonic and trionic spectra}. 

Analyzing the calculated absorption spectra, we further reveal the existence of higher energy interlayer excitons with a larger delocalization in reciprocal space and high binding energy, which can be interpreted as interlayer analog to the prominent C excitons in TMDCs mono- and few-layer materials.

\section{Acknowledgements}
Computational resources used for the simulations in this work were provided by the North-German Supercomputing Alliance (HLRN) under Project bep00047 and by the Regional Computing Center Erlangen (RRZE). 


\end{document}